\documentclass[12pt,a4]{article}
\usepackage{amsfonts}
\usepackage[latin1]{inputenc}
\usepackage[all]{xy}
\newcommand{\al}{{\alpha}}

\renewcommand{\th}{\vartheta}

\newcommand{\om}{{\omega}}

\newcommand{\Om}{{\Omega}}
\def\lp{\left(}
\def\rp{\right)}
\def\l[{\left[}
\def\r]{\right]}
\newcommand{\tiodot}{{\mbox{\footnotesize$\,\odot \,$}}}
\newcommand{\ticup}{{\mbox{\footnotesize$\,\cup \,$}}}
\newcommand{\teger}[1]{{\stackrel{\mathbb{Z}}{#1}}}

\newcommand{\prt}{\partial}
\newcommand{\tch}{{$\check{\mbox{C}}$}ech}
\newcommand{\BD}{{Deligne-Beilinson}}
\newcommand{\td}{\tilde{d}}
\newcommand{\cg}{\stackrel{\mathbb{Z}}{c}}
\def\ba{\begin{array}}
\def\ea{\end{array}}
\def\beq{\begin{equation}}
\def\eeq{\end{equation}}
\def\bea{\begin{eqnarray}}
\def\eea{\end{eqnarray}}

\def\om{\omega}
\def\Om{\Omega}

\newcommand{\ie}{{\em {i.e. }}}

\begin{document}

\newcommand{\cD}{{\cal{D}}}
\newcommand{\ZZ}{{\mathbb Z}}
\newcommand{\egzz}{\, \stackrel{\mathbb{Z}}{{=}}\,}
\newcommand{\egz}{\,{=}_{\mathbb{Z}}\,}
\pagestyle{empty} \setcounter{page}{0}
%% EQUATIONS AND DELIMITERS %%%
%%%%%%%%%%%%%%%%%%%%%%%%%%%%%%%

\newcommand{\sect}[1]{\setcounter{equation}{0}\section{#1}}
\renewcommand{\theequation}{\thesection.\arabic{equation}}
%\renewcommand{\theequation}{\thesubsection.\arabic{equation}}
%\renewcommand{\thesection}{\Roman{section}}
%\renewcommand{\thesubsection}{\thesection.\arabic{subsection}}

%%%%%%%%%%%%%%%%%%%%%%%%%%%%%%%%%%%%%%%%%%%%%%%%%%%%%%%%%%%%%%
%%%%%%%%%%%%%%%%%%%% LOGO LAPTH - DEBUT  %%%%%%%%%%%%%%%%%%%%
%%%%%%%%%%%%%%%%%%%%%%%%%%%%%%%%%%%%%%%%%%%%%%%%%%%%%%%%%%%%%%%
%{\Large{\bf {  {\today}  }}} \\

\newcommand{\LAP}{LAPTH}
\def\logo{{\bf {\huge LAPTH}}}
\centerline{\logo}
\vspace {3mm}
\centerline{{\bf{\it\Large Laboratoire d'Annecy-le-Vieux de
Physique Th\'eorique}}}
\centerline{\rule{12cm}{.42mm}}
%%%%%%%%%%%%%%%%%%%%%%%%%%%%%%%%%%%%%%%%%%%%%%%%%%%%%%%%%%%%%%%
%%%%%%%%%%%%%%%%%%%%% LOGO LAPTH  - FIN %%%%%%%%%%%%%%%%%%%%%
%%%%%%%%%%%%%%%%%%%%%%%%%%%%%%%%%%%%%%%%%%%%%%%%%%%%%%%%%%%%%%%

\vspace{20mm}

\begin{center}
{\Large {\bf A CLASS OF TOPOLOGICAL ACTIONS}}%
\\[1cm]

{\large M. Bauer$^{a}$, G. Girardi$^{b}$, R. Stora$^{b}$ and F.
Thuillier$^{b}$}

\end{center}

$^{a}$ {\it Service de Physique Théorique de Saclay, F-91191,
Gif-sur-Yvette, France}.

$^{b}$ {\it LAPTH, Chemin de Bellevue, BP 110, F-74941
Annecy-le-Vieux cedex, France}.

\vspace{20mm}

\centerline{{\bf Abstract}}
%\begin{center}
We review definitions of generalized parallel transports in terms of
Cheeger-Simons differential characters. Integration formulae are
given in terms of {\BD} cohomology classes. These representations of
parallel transport can be  extended to situations involving
distributions  as is appropriate in the context of quantized fields.
%\end{center}

\vspace{5mm}

\indent

\vfill
%\rightline{hep-th-\0406221}
\rightline{\LAP-1099/05} \rightline{SPhT-T04/080}
\newpage
\pagestyle{plain} \renewcommand{\thefootnote}{\arabic{footnote}}
\newpage
\tableofcontents \newpage \pagestyle{plain}

\sect{Introduction}

\hspace{1.0cm}Parallel transports and generalizations thereof have
been repeatedly met both in mathematics
\cite{Gun66,CS73,K73/74,Del71,Del99,B85,Hop02} and in global
aspects of gauge theories \cite{Alv84,G87,Fr92}, which played a
major role in elementary particle physics.

\noindent It has taken some time for the existing mathematics
\cite{EV88,J88,Bry93,Hop02,HLZ03,A02} to become known to
physicists \cite{G87,AT00, Fr92,AO00,Zuc00,AO03}.

\noindent At the semi-classical level one is lead to integrate
objects more general than differential forms over cycles with a
result defined modulo integers; Cheeger-Simons differential
characters \cite{CS73} are privileged candidates. Their integral
representations in terms of {\BD} smooth cohomology classes are
particularly well adapted to field theory for two reasons : first
of all, they involve locally defined fields subject to some gluing
properties. Besides, they allow for natural generalizations well
adapted to, at least, semi-classical quantization. Indeed the
latter already requires regularizing (thickening) the integration
cycles, an operation which can be performed easily within the
{\BD} cohomology framework. This operation is less naive than one
might think; indeed the corresponding currents are not any longer
differential forms (de Rahm currents) but {\BD} classes. In view
of this phenomenon we shall proceed in detail from the
semi-classical situation for which the use of Cheeger-Simons
characters is well adapted. In this case there exist canonical
integral representations in terms of differential forms with
discontinuous coefficients and therefore inappropriate  for
applications to quantum fields, even in the semi-classical
approximation. Fortunately, these integral representations can be
replaced by others with smooth coefficients. The latter are easily
generalizable to situations involving distributions and therefore
well adapted to quantum fields. There is however a price to pay:
the differential forms involved in the classical formulae have to
be replaced (non canonically) by {\BD} smooth classes.

\noindent

We start in section 2 with the prototype example of Maxwell's
electromagnetism in which a functional integral is defined under
"reasonable" hypotheses concerning the interaction with an
external current. The rest of the paper is devoted to a sequence
of constructions which give a mathematical foundation of the above
hypotheses.

 \indent Section 3 proposes three equivalent ways to describe Cheeger-Simons
 differential characters in terms of the integration of {\BD} cohomology classes.

 \indent Section 4 presents the natural generalizations required upon
quantization: the integration of {\BD} classes with distributional
coefficients.

 \indent Section 5 contains our  concluding remarks.

  \indent A number of technical details are collected in three appendices.

\sect{Maxwell Semi-Classical theory (à la Feynman)}
\label{sec:example} \hspace{1.0cm}
 While in a classical theory the action (when it exists) is
optional (in principle, the equations of motion are sufficient),
it becomes the keystone of the Feynman semi-classical point of
view. Hence, such an action must be carefully defined. In the
context of Maxwell's electromagnetism, we consider the Euclidean
action defined on a 4-dimensional, riemmannian \footnote{* is the
usual Hodge operator.}, compact manifold $M_4$
\begin{equation}\label{a5}
S_{EM}=  \frac{1}{2} \,{}\int_{{M}_4}{F \wedge *F} \, + \, i \cdot
``\int_{{M}_4}{j \wedge A}\," \,.
\end{equation}
\noindent Quotes emphasize that we have to make precise the
meaning of the second integral since $A$ is {\bf{not}} a 1-form on
$M_4$, but rather a connection on a $U(1)$-bundle over $M_4$, with
curvature $F$. We defer until the next section a mathematically
sound definition \footnote{It will turn out that more data than
just the 3-form $j$ will be needed.} of $"\int_{{M}_4}{j \wedge
A}\,"$ for $j$ a 3-form with integral periods.

 At this point, we
only need to know that $"\int_{{M}_4}{j \wedge A}\,"$ will be
defined modulo $2\pi \mathbb{Z}$ and will fulfill the following
natural property : if $A = A_0 + \alpha$ (with $A_0$ a fixed
$U(1)$-connection and $\alpha$ a generic 1-form), then

\begin{equation}\label{a6}
  ``\int_{{M}_4}{j \wedge A}\," = ``\int_{{M}_4}{j \wedge A_0}\," + \int_{{M}_4}j
\wedge \alpha .
\end{equation}
 Gauge invariance requires $\int_{{M}_4}j
\wedge (g^{-1}dg) \in 2\pi \mathbb{Z} \,$ which is less
restrictive than the "classical" requirement $\int_{{M}_4}j \wedge
(g^{-1}dg) = 0$, commonly assumed \cite{W99,Zuc03} to hold at the
quantum level.

\indent Once the choice of definition of the action integral with
the above property has been made, we can try to evaluate the state
\footnote{A linear functional on observables.} ($\hbar=1$)
\begin{equation}
<{\mathit   e}^{ - \ i \cdot \, ``\int_{{M}_4}{j \wedge A}\,"}> \
= \int {\cD} A \,{\mathit   e}^{- \frac{1}{2} \int_{{M}_4} F
\wedge *F \, - \, i \cdot \, ``\int_{{M}_4}{j \wedge A}\,"},
  \label{3}
\end{equation}
where $A$ is a $U(1)$-connection. First let
\begin{equation}
A = A_0 + \alpha,
  \label{4}
\end{equation}
with $A_0$ a background connection and $\alpha $ a globally
defined 1-form. Then, denoting by  $F_0 = d A_0$ the background
curvature, we obtain
\begin{eqnarray}
< {\mathit   e}^{ - \ i \cdot \, ``\int_{{M}_4}{j \wedge A}\,"} >
&=& {\mathit e}^{- \frac{1}{2} \int_{{M}_4} F_0 \wedge *F_0 \, -
\, i \cdot \, ``\int_{{M}_4}{j \wedge A_0}\,"}\nonumber \\  &&
\hspace{-3mm} \times \int {\cal{D}} \alpha \,{\mathit   e}^{-
\frac{1}{2} \int_{{M}_4}{d \alpha \wedge *d \alpha} \, - \,
\int_{{M}_4}{F_0 \wedge *d \alpha} \, - \, i \cdot \,
\int_{{M}_4}{j \wedge \alpha} }.
  \label{5}
\end{eqnarray}
The 1-form $\alpha$ is linearly coupled to $(j + i\, d*F_0)$ and
we need to gauge fix the $\alpha$ integration. Note that $\int j
\wedge \alpha$ is an ordinary integral.
 Gauge transformations connected with the identity are eliminated by
choosing a Green function ($\xi$, the gauge parameter)
\begin{equation}
G_\xi \, = \, [ \delta d + \xi d \delta ]^{-1} \ \  ,\ \xi >0,
  \label{6}
\end{equation}
in the subspace orthogonal to harmonic forms (the elimination of
large gauge transformations will come later). So, we are led to
\begin{eqnarray}
\hspace{-13mm}<{\mathit e}^{- \ i \cdot \, ``\int_{{M}_4}{j \wedge
A}\,"}> &=& {\mathit   e}^{- \frac{1}{2} \int_{{M}_4} F_0
  \wedge *F_0 \, - \, i \cdot \,
``\int_{{M}_4}{j \wedge A_0}"} \nonumber \\
&& \hspace{-3mm}\times {\mathit   e}^{- \frac{1}{2} \int_{{M}_4}
(j + i\,d * F_0)_\perp
  G_\xi \, *(j+i \, d * F_0)_\perp} \times Z (j_\parallel).
  \label{7}
\end{eqnarray}
The subscript $\perp$ (resp. $\parallel$) refers to the
decomposition of forms into components orthogonal to (resp. along)
harmonic forms. We shall come to the definition of
$Z(j_\parallel)$ later.

The $A_0$ dependence can be reduced to:
\begin{eqnarray}
\hspace{-13mm}<{\mathit   e}^{- \ i \cdot \, ``\int_{{M}_4}{j
\wedge A}\,"}> &=& {\mathit   e}^{- \frac{1}{2} \int_{{M}_4}
F_{0\parallel}
\wedge * F_0 \ - \ i \cdot \, ``\int_{{M}_4}{j_\parallel \wedge A_0}"} \nonumber \\
&& \hspace{-3mm}\times {\mathit   e}^{- \frac{1}{2} \int_{{M}_4}
j_\perp G_\xi \, * j_\perp} \cdot Z (j_\parallel).
  \label{8}
\end{eqnarray}
The first term yields an overall normalization factor to be
divided out. The third term is $\xi$ independent by $dj=0$. The
forms $\alpha_\parallel$ and $j_\parallel$ being harmonic are
necessarily closed (also co-closed). Using Poincaré duality
 and assuming no torsion, we can
decompose them along a dual basis of integral 3-cycles and
1-cycles respectively
\begin{eqnarray}
\alpha_{\parallel} = \sum_k \alpha_k \,{{\zeta}_k}^{(3)}
+ d ( \cdots ) \ \   &,&\ \ j_\parallel = \sum_k n_k {\zeta_k}^{(1)}
+ d ( \cdots )\nonumber \\
 \ \ \
with \ \  {<{\zeta}_k}^{(3 )}\!\!\!\!&,&\!\!{\zeta_l}^{(1)}>\  =\
\delta_{kl},
  \label{10}
\end{eqnarray}
where the $\alpha_k$'s are real numbers since $\alpha$ is real,
while the $n_k$'s are integers since $j_\parallel$ has integral
periods. With this decomposition of $\alpha_\parallel$ and
$j_\parallel$, we can formally write
\begin{equation}
Z (j_\parallel) = \int {\cal{D}} \alpha_\parallel \,{\mathit e}^{
\, i \cdot \int_{{M}_4}{\alpha_\parallel \wedge j_\parallel}} =
\int \ d \vec{\alpha} \,{\mathit e}^{ \, i \vec{n}\cdot
\vec{\alpha}} \ , \label{9}
\end{equation}
where $\vec{\alpha} = ( \alpha_1,...,\alpha_m) $ and $\vec{n} =
(n_1,...,n_m)$.

Now, large gauge transformations are :
\begin{equation}
\alpha_k \mapsto \alpha_k + p_k, \ \  p_k \in 2\pi\mathbb{Z}
  \label{11}
\end{equation}
and can be factored out by transforming $ \alpha_k$ integration
into $\vartheta_k$ integration $0 \leq \vartheta_k < 2 \pi$:
\begin{equation}
Z (j_\parallel) = \int   d \vec{\vartheta} \,{\mathit   e}^{ \, i
\vec{n} \cdot \vec{\vartheta}}.
  \label{12}
\end{equation}
These angles $\vartheta_k$ parametrize
$H^1({{M}_4},\mathbb{R})/H^1({{M}_4},\mathbb{Z})$, still assuming
no torsion (torsion yields an extra factor).

Similarly
\begin{equation}
{\mathit   e}^{ \, i \cdot \, ``\int_{{M}_4}{j_\parallel \wedge
A_0}"} =
 {\mathit   e}^{ \, i \vec{n} \cdot \vec{\vartheta}_{0}},
  \label{13}
\end{equation}
where $\vartheta_{0k}$ are fixed angles which may be incorporated
into $\vartheta_k$.

To conclude, after normalization, the state $< \ \ >$ can be
decomposed into gauge invariant states labelled by the angles
$\vec{\vartheta}$
\begin{equation}
<{\mathit   e}^{ \, i \cdot \, ``\int_{{M}_4}{j \wedge A}\,"}> \ =
\int d \vec{\vartheta} <{\mathit e}^{ \, i \cdot \,
``\int_{{M}_4}{j \wedge A}\,"}>_{\vec{\vartheta}} \ ,
  \label{14}
\end{equation}
with
\begin{equation}
<{\mathit   e}^{ \, i \cdot \, ``\int_{{M}_4}{j \wedge
A}\,"}>_{\vec{\vartheta}} \ = {\mathit   e} ^{ \, i  \vec{n} \cdot
\vec{\vartheta}} \cdot {\mathit e}^{-\frac{1}{2} \int_{{M}_4}
  j_\perp G_\xi \, *j_\perp},
  \label{15}
\end{equation}
a familiar situation which provides an alternative to the commonly
accepted choice \cite{W99,Zuc03} which amounts to integrate over
$\vec{\vartheta}$'s with the result $\propto \delta
(j_\parallel)$; in the latter case $j=dm$ are the only possible
integration currents for $A$, while for the states defined in
(\ref{15}) the currents $j$ are only required to be closed forms
with integral periods. In other words, homological triviality of
Wilson loops or appropriately smeared version thereof are not
consequences of gauge invariance, but rather, of some form of
locality.

\vspace{5mm}

\sect{Integral representations of differential characters}
\label{sec:repres}
\hspace{1.0cm} In section \ref{sec:example} we
have described the physical consequences of `` $\int_M{j \wedge
A}$\, " being defined modulo $2\pi \mathbb{Z}$ (with $j$ a form
with integral periods). We shall now proceed to give some
substance to this assumption and write down explicit formulae.

To start with, let us recall that one can associate to any closed
curve\footnote{By {\em curve} we mean a 1-dimensional embedded
smooth submanifold of $M$.} $\Gamma$ in a manifold $M$ a closed
current $\delta_\Gamma$ ($i.e.$ a closed form whose local
representatives have distributional coefficients) such that
integration of a form $\omega$ along $\Gamma$ formally reduces to
the integration of $\delta_\Gamma \wedge \omega$ over the whole of
$M$ \cite{dR55}.

We shall first try to find a satisfactory definition of the {\em
circulation integral} of $A$ along a closed curve $\Gamma$ by
considering various situations. This study will naturally lead us
to the mathematical notion of {\em differential character}
introduced by Cheeger and Simons \cite{CS73}.

Then, while seeking for a representation of a differential
character supported by {\tch}-de Rham cohomology theory, there
will emerge a  defining formula for `` $\int_M{j \wedge A}$\, " in
terms of {\BD} cohomology \cite{Bry93}. We will see that for
$j=\delta_{\Gamma}$, there is a canonical definition of this
integral, whereas for general $j$ there is a whole class of
adequate definitions.

 From now on, $M$ will be a \textbf{torsion-free} smooth
$n$-dimensional oriented compact manifold without boundary.

\subsection{Circulation of $U(1)$ gauge fields as
differential characters} \label{subsec:chardiff}
 \hspace{1.0cm}Within Maxwell's theory of electromagnetism on ${{M}_4}$, due to
the triviality of the homology and cohomology groups of ${\mathbb
R}^4$ ($i.e.$ any closed curve is a boundary, and any closed
3-form is exact), the circulation of a $U(1)$-gauge field $A$
along a closed curve $\Gamma $ is a perfectly well-defined and
gauge invariant integral which measures the magnetic flux through
any surface $\Sigma$ with boundary $\Gamma= \prt \Sigma$, namely
\begin{equation} \label{stok}
``\oint_{\Gamma} A \," \equiv \oint_{\Gamma= \,\prt \Sigma} A \, =
\, \int_{\Sigma}F.
\end{equation}
Of course, such a property fails for a general manifold $M$ with
non-trivial (co-)homology groups. Nevertheless, it may be asked
whether (\ref{stok}) can be maintained for boundaries $\Gamma=
\prt \Sigma$, assuming that ``$\oint_{\Gamma} A$" has a
mathematical meaning for any closed curve $\Gamma$ in $M$. Let us
then consider a closed curve $\Gamma $ that splits a {\bf closed}
surface $\Sigma $ into two components $\Sigma_+$ and $\Sigma_-$ :
$\Sigma = \Sigma_+ \cup \Sigma_-$ and $\Gamma= \prt \Sigma_+ =
-\prt \Sigma_- $, where the minus sign takes care of orientations.
Then, we would have
\begin{equation} \label{circ+}
``\oint_{\Gamma} A \," =\,\int_{\Sigma_+}F
\end{equation}
since $\Gamma= \prt \Sigma_+$, and
\begin{equation} \label{circ-}
``\oint_{\Gamma} A \," = \, -\int_{\Sigma_-}F
\end{equation}
since $\Gamma= \prt \Sigma_-$. Since $F$ is a $U(1)$ curvature, we
know that
\begin{equation} \label{Fchern}
\oint_{\Sigma_-} F \, + \, \oint_{\Sigma_+} F \, = \,
\oint_{\Sigma} F \in \mathbb{Z}(1) := 2i \pi \mathbb{Z}
\end{equation}
on any closed surface $\Sigma$. This suggests that, if it exists,
``$\oint_{\Gamma} A$" is only defined modulo $\mathbb{Z}(1):=2i\pi
\mathbb{Z}$. Otherwise stated, we can expect, for fixed $A$,
``$\frac{1}{2i\pi}\oint_{\Gamma} A$" to be some
$\mathbb{R}$/$\mathbb{Z}$-valued linear functional on the space of
closed curves (cycles). Let us have a closer look at such an
assumption.

To begin with, a $U(1)$-gauge transformation, $g$, changes the
connection $A$ into the connection $A^g = A + g^{-1}dg$ with the
same curvature $F$; therefore, {\bf{if (\ref{stok}) holds}}
\begin{equation} \label{circinv}
``\oint_{\prt \Sigma} A^g \," =``\oint_{\prt \Sigma} A
+g^{-1}dg\," =\,\int_{\Sigma}F\, = \, ``\oint_{\prt \Sigma} A \,"
,
\end{equation}
{\ie}``$\oint_{\prt \Sigma} A$"  is gauge invariant.

In fact, for any closed $1$-form $\alpha$ on $M$, $A + \alpha$ is
also a connection with curvature $F~=~dA$, so that we obtain a
relation similar to (\ref{circinv})  with $\alpha$ in place of $
g^{-1}dg$. Consequently, we can infer that connections with the
same curvature may define (\emph{a priori} different)
$\mathbb{R}$/$\mathbb{Z}$-valued linear functionals on cycles
which \emph{coincide on boundaries}. In this sense, the
``integral" of $A$ on boundaries is completely defined by $F$.

For a general closed curve $\Gamma$ and any gauge transformation
$g$, we would like to maintain gauge invariance of
$``\oint_{\Gamma} A \ "$, which is not immediate since the term
$\oint_{\Gamma} g^{-1}dg$  may not vanish ($\Gamma$ not being
necessarily a boundary). However, since $g^{-1}dg$ is the pullback
by $g$ of the standard $U(1)$ ($\simeq S^1$) volume $1$-form,
$z^{-1}dz$, we have
\[
\oint_{\Gamma} g^{-1}dg \in \mathbb{Z}(1) \ .
\]
Accordingly, still assuming that $``\oint_{\Gamma} A \,"$ is
defined modulo $\mathbb{Z}(1)$, we obtain the sought after gauge
invariance
\begin{equation} \label{gaugegal} ``\oint_{\Gamma} A^g
\," = \, ``\oint_{\Gamma} A \," \,,
\end{equation}
 though $\Gamma$ is {\bf{not}} a boundary.

All these requirements can be satisfied if we ask for (\ref{stok})
and define $``\frac{1}{2i\pi}\oint_{\Gamma} A "$ to be an
$\mathbb{R}$/$\mathbb{Z}$-valued functional, linear in $\Gamma$
and affine in $A$ {\footnote{This is a natural demand since the
space of connections is an affine space.}}, a property which is
satisfied if we set
\begin{equation}\label{affin}
``\oint_{\Gamma} (A + \gamma)" = ``\oint_{\Gamma} A "
+\oint_\Gamma \gamma,
\end{equation}
where the last integral is the ordinary integral of the 1-form
$\gamma$ -in the same line of thought recall (\ref{a6})-. Then,
for any closed $1$-form $\al$ we have
\begin{equation}\label{invA}
``\oint_{\Gamma} (A + \al)" = ``\oint_{\Gamma} A " \
\end{equation}
if and only if all periods of $\al$ take values in
$\mathbb{Z}(1)$. In fact the $1$-forms $g^{-1}dg$, with $g$
running through the $U(1)$-gauge group, generate the space of
closed $1$-forms with $\mathbb{Z}(1)$-valued periods. That is, if
$per(\al) \in \mathbb{Z}(1)$, we can write \[\al \, = \, g^{-1}dg
+ d \lambda\] for some $U(1)$-gauge transformation $g$ and some
function $\lambda$ on $M$. Then, as far as ``integration" of $A$
on closed curves is concerned, gauge invariance is equivalent to
invariance under $A \, \mapsto \, A + \al$, with $\al$ a form with
$\mathbb{Z}(1)$-valued periods. Therefore, it is expected that two
connections that differ by a form with $\mathbb{Z}(1)$-valued
periods define the same $\mathbb{R}$/$\mathbb{Z}$-valued linear
functional on the space of closed curves.

At this point, let us make some remarks.
 First, if the connection $A$ is a $1$-form on $M$ (for instance when
the corresponding $U(1)$-bundle is flat), we must require that the
general definition of $``\oint_{\Gamma} A \,"$ reduces to the
usual definition of the integral of a form. Second, up to now, we
have only considered $U(1)$-connections on $M$. In a more general
situation we will consider objects $A^{(p)}$, representing
antisymmetric tensor ``gauge potentials"  which appear in
supergravities and string theories \cite{Jo02}. However, the
geometric situation turns out to be more involved than in the case
of connections. Indeed, a $U(1)$-connection, although it is not a
$1$-form on $M$, is lifted as a $1$-form on some principal
$U(1)$-bundle over $M$. Such $A^{(p)}$'s will in general not be
$p$-forms on $M$. It turns out that they can be considered as
connections on new mathematical objects called gerbes
\cite{Bry93,Zu00}. Here we will not go into such an
interpretation: we will consider locally defined differential
forms $``A^{(p)}"$ on $M$ whose differentials, $F^{(p+1)}$, are
globally defined $(p+1)$-forms with $\mathbb{Z}$-valued periods on
$M$ \footnote{In this framework, $A =(2i \pi)A^{(1)}$, and its
curvature $F=(2i \pi)F^{(2)}$.}. We will define an
$\mathbb{R}$/$\mathbb{Z}$-valued linear functional,$``\oint_{S_p}
A^{(p)}"$  on the space of closed $p$-submanifolds, $S_p$, of $M$.
Such linear functionals turn out to be {\em differential
characters} in the sense of J. Cheeger and J. Simons. Differential
characters have been constructed within the framework of
Chern-Simons' theory of secondary characteristic classes, an
extension of the Chern-Weil theory. They were introduced to
describe, on the base space, secondary characteristic classes of
principal bundles initially defined as differential forms on the
whole bundle space (see \cite{K73/74} for a review, and
\cite{CS73} for the original reference).

Our integrals, $``\oint_{S_p}A^{(p)}"$, are related to {\em{{\BD}
cohomology classes}} as presented in \cite{Bry93} and therefore
(cf. section 7 of appendix A) offer a parametrization of
differential characters.

In appendix \ref{appDB} the reader will find notations, basic
definitions and results concerning smooth {\BD} cohomology groups
$H^{q}({\cal{C}}_{p},D)$.

Our basic example deals with a $U(1)$-connection on the $n$-
dimensional manifold $M$. In this case there is a one to one
correspondence between  the second smooth Deligne cohomology group
of $M$, $H^2({\cal{C}}_2,D)$ \footnote{cf. appendix A.}, and the
set of equivalence classes of $U(1)$ principal bundles with
connection, $(P[U(1)],A)$ (cf. appendix \ref{app3}). We will show
how to integrate an element of $H^2({\cal{C}}_2,D)$ over a
1-cycle, $z_1$, and take this ``integral" as a definition for
$``\frac{1}{2i\pi}\oint_{\Gamma} A "$. This generalizes to
integrating elements of $H^{p+1}({\cal{C}}_{p+1},D)$ over
$p$-cycles, $z_p$ which provides a definition for
$``\oint_{z_p}A^{(p)}"$. As we shall see (section
\ref{subsec:Cycledef}) the classical Weil construction, pertaining
to singular homology,  both suggests a natural definition of
elements of $H^{p+1}({\cal{C}}_{p+1},D)$ and of their integration
over a $p$-cycle. In \cite{Zuc00} R. Zucchini gives integral
representations of ``relative" differential characters,
essentially identical with ours, independently of the expression
of the integrand in terms of {\BD} classes. Later (section
\ref{subsec:longf}) we will give another definition of the
integral which avoids Weil's analysis of the cycle and allows for
generalization.

\subsection{Integration over a cycle: the appearance of {\BD}
classes} \label{subsec:Cycledef}
 \hspace{1.0cm} There is a natural
procedure to define integration over integral cycles, based on the
classic work of André Weil \cite{AW52}. In this paper, for any
simple \footnote{Definitions and notations are given in appendix
\ref{appDB}.} covering ${\cal{U}}$ of $M$,
"${\cal{U}}$-$p$-chains" are defined as singular $p$-chains,
$C_p$, such that
\begin{equation}\label{de cadix}
  C_p  = {\partial} C_{(0,p)} :=\sum_\al C_{(0,p),\, \al},
\end{equation}
where every $C_{(0,p),\, \al}$ is a singular $p$-chain with
carrier $\mathcal{U}_\al$ (here, $\partial$ is the boundary
operator on {\tch} chains). A ${\cal{U}}$-$p$-cycle $z_p$ is a
closed ${\cal{U}}$-$p$-chain ($bz_p=0$, with $b$ the boundary
operator on singular chains). Then, it is shown that for any
${\cal{U}}$-$p$-cycle $z_p$ of $M$ there exists a sequence of
{\tch} (smooth) singular ${\cal{U}}$-chains, $z_{(k,p-k)}$
\begin{eqnarray}\label{Wcycle}
z^{\mathcal{W}}_{(p)} := ( z_{(0,p)} , \ldots , z_{(k,p-k)} ,
\ldots ,z_{(p,0)} ) ,
\end{eqnarray} where each $z_{(k,p-k)}$ has support in some open  $(k+1)$-fold intersection of
${\cal{U}}$, such that
\begin{eqnarray}\label{Wdescent1}
\partial z_{(0,p)} &=& z_{(-1,p)} := z_p  \nonumber\\
bz_{(k,p-k)} &=& \partial z_{(k+1,p-k-1)}, \ \ k \in \{ 1,...,p-1
\} \nonumber
\\b_0z_{(p,0)} &:=& z_{(p,-1)}  ,
\end{eqnarray}
where $b_0$ is just the ``degree" operator on singular chains
\cite{AW52}, $z_{(p,-1)}$ is an integral {\tch} $p$-cycle of
${\cal{U}}$ and $(\prt z_{(k,p-k)})_{\al_0
\cdots\al_{k-1}}=\sum_{\beta}\, z_{(k,p-k),{\beta\al_0
\cdots\al_{k-1}}}$.

 The collection
$z^{\mathcal{W}}_{(p)}$ is called a Weil descent of $z_p$, and the
corresponding equations (\ref{Wdescent1})  a Weil descent equation
of $z^{\mathcal{W}}_{(p)}$.

Now, if $Z^{\mathcal{W}}_{(p)}$ is another Weil descent of the
same ${\cal{U}}$-$p$-cycle $z_p$, it differs from
$z^{\mathcal{W}}_{(p)}$ according to
\begin{eqnarray}\label{Wambiguity}
Z_{(0,p)} &=& z_{(0,p)} +\partial t_{(1,p)}+b t_{(0, p+1)},
\nonumber  \\ Z_{(k,p-k)} &=& z_{(k,p-k)} +bt_{(k,p-k+1)} +
\partial t_{(k+1,p-k)}\ \ ,\ \ k = 1,...,p-1  \nonumber \\
Z_{(p,0)} &=& z_{(p,0)} + bt_{(p,1)} + \partial t_{(p+1,0)} ,
\end{eqnarray}
where the $t_{(k,p-k+1)}$ are some {\tch} ${\cal{U}}$-chains.
Since $z_p$ is fixed, we must have
\begin{equation}
\partial  b\, t_{(0,p+1)} =\, 0 \, = b  \partial \, t_{(0,p+1)},
\end{equation}
which means that $\partial t_{(0,p+1)} $ is a
${\cal{U}}$-$(p+1)$-cycle, $\tilde{z}_{p+1}$ which in turn gives
rise to a Weil descent $$\tilde{z}^{\,\mathcal{W}}_{(p+1)} :=
(\tilde z_{(0,p+1)} := t_{(0,p+1)} ,\tilde z_{(1,p)} , \ldots ,
\tilde z_{(k,p-k+1)} , \ldots,\tilde z_{(p+1,0)} ),$$ so that
\begin{eqnarray}
Z_{(0,p)} &=& z_{(0,p)} +\partial ( t_{(1,p)}+{\tilde z}_{(1,p)}),\nonumber\\
 Z_{(k,p-k)} &=& z_{(k,p-k)} +b( t_{(k,p-k+1)}+{\tilde
z}_{(k,p-k+1)}) + \partial
(t_{(k+1,p-k)}+ {\tilde z}_{(k+1,p-k)}), \,\  \nonumber \\
Z_{(p,0)} &=& z_{(p,0)} + b(t_{(p,1)} +{\tilde z}_{(p,1)}) +
\partial ( t_{(p+1,0)}+ {\tilde z}_{(p+1,0)}) ,
\end{eqnarray}
with $k = 1,...,p-1 $. Accordingly, the general ambiguities on a
Weil descent of a given cycle $z_p$ of $M$ take the form
\begin{eqnarray}\label{BDambiguity}
Z_{(0,p)} &=& z_{(0,p)} +\partial h_{(1,p)},\nonumber\\
 Z_{(k,p-k)} &=& z_{(k,p-k)} +b h_{(k,p-k+1)} + \partial
h_{(k+1,p-k)}, \,\  \nonumber \\
Z_{(p,0)} &=& z_{(p,0)} + bh_{(p,1)} +
\partial h_{(p+1,0)} ,
\end{eqnarray}
By identifying Weil descents that differ by ambiguities
(\ref{BDambiguity}), one defines an equivalence relation between
Weil descents whose corresponding equivalence classes
\emph{canonically} represent ${\cal{U}}$-$p$-cycles of $M$.
Actually, one could introduce a boundary operator made of the
operators $b$ and $\partial$, turning what we have just done into
a homological game in which Weil descent classes are homology
classes.

Similarly -cf. appendix \ref{appDB}-, a sequence
\begin{eqnarray}\label{BDcocycle}
\omega_{\mathcal{D}}^{(p)} := (\omega^{(0,p)} , \omega^{(1,p-1)} ,
\ldots , \omega^{(p,0)} , {\mathop \omega
\limits^{\mathbb{Z}}{^{(p+1,-1)}}} ),
\end{eqnarray}
where $ \om^{(k,p-k)} \in {\check{\mbox{C}}}^k
\left({\cal{U}},\Omega^{(p-k)}(M)\right) $ and ${\mathop \omega
\limits^{\mathbb{Z}}{^{(p+1,-1)}}} \in {\check{\mbox{C}}}^{(p+1)}
({\cal{U}})$ defines a {\BD} cocycle if
$$(\tilde d+\delta)\omega_{\mathcal{D}}^{(p)} = D\omega_{\mathcal{D}}^{(p)} =0,$${\ie}
\begin{equation}
  d_{p-k} {\om}^{(k, p-k)} \  =\ \delta {\om}^{(k-1, p-k+1)},\ \ \ k=1,\cdots ,p+1\ .
\end{equation}
In the above equation $\delta$  is the {\tch} coboundary operator,
$d_{-1} {\mathop \omega \limits^{\mathbb{Z}}{^{(p+1,-1)}}}$ is the
injection of numbers into $\Omega^{(0)}(M)$ and $\tilde d$ the
differential of the Deligne complex (it coincides with the de Rham
differential $d$, up to degree $p-1$ and is the \textbf{zero} map
at degree $p$). By convention, cohomology (resp. homology) indices
are upper (resp. lower) indices, those referring to {\tch} complex
coming first.

Note that ${\mathop \omega \limits^{\mathbb{Z}}{^{(p+1,-1)}}}$ is
necessarily a cocycle, and, although $\tilde d \om^{(0,p)} \equiv
0$, $ d \om^{(0,p)}$ is the restriction of a globally defined
closed form $\om^{(-1,p+1)}$ with integral periods \cite{AW52}.
This $\om^{(-1,p+1)}$ will be called the top form of the cocycle
$\omega_{\mathcal{D}}^{(p)}$.

We can now proceed and build {\BD} cohomology classes as
equivalence classes of {\BD} cocycles related as follows:
$$\varpi_{\mathcal{D}}^{(p)}=\omega_{\mathcal{D}}^{(p)}
+D\,Q_{\mathcal{D}},$$ {\ie}
\begin{eqnarray}\label{Bambiguity}
\varpi^{(0,p)} &=& \om^{(0,p)} + d q^{(0,p-1)}, \nonumber \\
\varpi^{(k,p-k)} &=& \om^{(k,p-k)} +d q^{(k,p-k-1)} +
\delta q^{(k-1,p-k)}\ \ ,\ \ k =1,...,p , \nonumber \\
 {\mathop \varpi \limits^{\mathbb{Z}}{^{(p+1,-1)}}} &=&
 {\mathop \omega \limits^{\mathbb{Z}}{^{(p+1,-1)}}} + \delta
  {\mathop q\limits^{\mathbb{Z}}{^{(p,-1)}}},
\end{eqnarray}
where $q^{(k,p-k-1)} \in {\check{\mbox{C}}}^k
\left({\cal{U}},\Omega^{(p-k-1)}(M)\right)$ and ${\mathop
q\limits^{\mathbb{Z}}{^{(p,-1)}}} \in {\check{\mbox{C}}}^p
\left({\cal{U}} \right)$. As an immediate consequence, all
cocycles belonging to the same {\BD} cohomology class have the
same top form.

The integral of a {\BD} cocycle $\omega_{\mathcal{D}}^{(p)}$ over
a $p$-cycle $z^{\mathcal{W}}_{(p)}$ is naturally defined as the
pairing
\begin{eqnarray}\label{integomega}
\int_{z^{\mathcal{W}}_{(p)}} \omega_{\mathcal{D}}^{(p)} \ &:=& \
<\omega_{\mathcal{D}}^{(p)},z^{\mathcal{W}}_{(p)}> \ := \
\sum\limits_{k = 0}^p \int_{z_{(k,p-k)}} \omega^{(k,p-k)}
\nonumber \\ &:=&  \sum\limits_{k = 0}^p \frac{1}{(k+1)!} \sum
_{\al_0, \cdots,\al_k}\int_{z_{(k,p-k),{\al_0 \cdots\al_k}}}
\omega^{(k,p-k)}_{\al_0 \cdots\al_k}\ .
\end{eqnarray}

 In (\ref{integomega}) the
ambiguities on the representatives ${z^{\mathcal{W}}_{(p)}}$
(resp. $\omega_{\mathcal{D}}^{(p)}$)
 of $[z^{\mathcal{W}}_{(p)}]$ (resp. [$\omega_{\mathcal{D}}^{(p)}$]) generate terms of the form
\begin{eqnarray}\label{integamb}
\int_{h_{(p+1,0)}} d_{-1}({\mathop \omega \limits^{\mathbb
Z}{^{(p+1,-1)}}} + \delta {\mathop q \limits^{\mathbb
Z}{^{(p,-1)}}}) \ + \ \int_{z_{(p,0)}} d_{-1}{\mathop q
\limits^{\mathbb Z}{^{(p,-1)}}} \ .
\end{eqnarray}
These terms are necessarily integers since the chains and the
cochains appearing there are integers. In other words,
(\ref{integomega}) extends to classes as long as we work modulo
``integers". This also means that the duality so realized is over
$\mathbb{R}/{\mathbb{Z}}$, not $\mathbb{R}$, \ie of Pontrjagin
type. Actually, this is not totally surprising since a {\BD}
cohomology class defines a form up to a form with integral periods
(cf. appendix \ref{appDB}).

Many of the equalities we will encounter only hold true $mod\
{\mathbb{Z}}$, accordingly we shall use the notation $``\egzz"$ to
mean $``= \, ... \, mod\ {\mathbb{Z}}"$.

With all this information, we finally set
\begin{eqnarray}\label{integclassomega}
\int_{z_p} [\omega_{\mathcal{D}}^{(p)}] :=
\int_{[z^{\mathcal{W}}_{(p)} ]} [\omega_{\mathcal{D}}^{(p)}] \,
\egzz \, \sum\limits_{k = 0}^p \int_{z_{(k,p-k)}} \omega^{(k,p-k)}
\ ,
\end{eqnarray}
for any representative of $[\omega_{\mathcal{D}}^{(p)}]$ and
$[z^{\mathcal{W}}_{(p)}]$ to which we shall refer to
(\ref{integclassomega}) as the {\emph{``Defining
Formula"}}.\textit{}

Let us note that the linearity of (\ref{integclassomega}) with
respect to $z_p$ is clear since all descents are linear.

\vspace{.5cm}
\subsubsection{Examples} \label{illus}
Let us apply (\ref{integclassomega}) to two simple cases. First,
consider the situation where the cycle $z_p$ is a boundary:
$z_p=bc_{p+1}$. Due to the equivalence of singular and {\tch}
homologies,  any {\tch} $p$-cycle, $z_{(p,-1)}$, arising from the
descent of $z_p$, is a {\tch} boundary, \emph{i.e.}
\begin{eqnarray}
z_{(p,-1)} = \partial c_{(p+1,-1)}  \ ,
\end{eqnarray}
for some integral {\tch} chain $c_{(p+1,-1)}$. Then, the
corresponding descent has a representative of the form
\begin{eqnarray}
z^{\mathcal{W}}_{(p)} := (z_{(0,p)}=bc_{(0,p+1)} , 0 \ldots , 0 ,
\ldots , 0) ,
\end{eqnarray}
with $\partial c_{(0,p+1)} = c_{p+1}$. Accordingly, the integral
of $[\omega_{\mathcal{D}}^{(p)}]$ over this trivial cycle $z_p$
reads
\begin{eqnarray}
\int_{z_p} [\omega_{\mathcal{D}}^{(p)}] &\egzz&
\int_{bc_{(0,p+1)}} \omega^{(0,p)} \egzz \int_{c_{(0,p+1)}}
d\omega^{(0,p)} \egzz
\int_{c_{(0,p+1)}} \delta_{-1} \omega^{(-1,p+1)} \nonumber \\
&\egzz& \int_{\partial c_{(0,p+1)}} \omega^{(-1,p+1)} \egzz
\int_{c_{p+1}} \omega^{(-1,p+1)} \ .
\end{eqnarray}
This property is exactly what we were expecting when we considered
the integration of a $U(1)$-connection (cf the introduction to
this section).

Second, let us assume that the $(p+1)$-form associated to
$[\omega_{\mathcal{D}}^{(p)}]$ is exact. Then, it can be shown
that there is a  {\BD} representative
\begin{eqnarray}
\omega_{\mathcal{D}}^{(p)} := (\omega^{(0,p)} = \delta_{-1}
q^{(-1,p)} , 0 \ldots , 0 , \ldots , 0)
\end{eqnarray}
of $[\omega_{\mathcal{D}}^{(p)}]$, where $\omega^{(-1,p+1)} =
dq^{(-1,p)}$. The integration formula now reads
\begin{eqnarray}
\int_{z_p} [\omega_{\mathcal{D}}^{(p)}] &\egzz& \int_{z_{(0,p)}}
\omega^{(0,p)} \egzz \int_{z_{(0,p)}} \delta_{-1} q^{(-1,p)} \nonumber\\
&\egzz& \int_{\partial z_{(0,p)}} q^{(-1,p)} \egzz \int_{z_p}
q^{(-1,p)} \ ,
\end{eqnarray}
as expected. Indeed, on the one hand, as we write
$\omega^{(-1,p+1)} = dq^{(-1,p)}$, we canonically associate to
$\omega^{(-1,p+1)}$ a definite form, on the other hand, we have
emphasized the fact that a {\BD} cohomology class
$[\omega_{\mathcal{D}}^{(p)}]$ defines a $p$-form on $M$, up to
$p$-forms with integral periods, $q^{(-1,p)}$. It is then natural
to find that the integral of $[\omega_{\mathcal{D}}^{(p)}]$ over a
cycle coincides -up to integers- with the integral of $q^{(-1,p)}$
over this cycle.

\subsection{An equivalent integration over the whole manifold}
\label{subsec:longf}
\hspace{1.0cm}In the previous approach that
led to the {\emph{Defining Formula}}, we  have only dealt with
integrals defined over cycles. In view of further generalization
we shall first express those as integrals over the whole manifold
$M$. A way to do so is to construct a version of Pontrjagin
duality in the {\BD} framework. In other words, we construct a
(non smooth) canonical {\BD} cohomology class
$[\eta_{\mathcal{D}}^{(n-p-1)}(z)]$ associated to any singular
$p$-cycle $z$ on $M$ and a cup product ($\cup_{\cal{D}}$)
\cite{B85,EV88, Bry93} such that
\begin{eqnarray}\label{Cyclethetaf}
\int_{z} [\omega_{\mathcal{D}}^{(p)}] &\egzz& \int_{M}
[\omega_{\mathcal{D}}^{(p)}] \cup_{\cal{D}}
[\eta_{\mathcal{D}}^{(n-p-1)}(z)] \ ,
\end{eqnarray} for any {\BD} cohomology class
$[\omega_{\mathcal{D}}^{(p)}]$. We refer the reader to appendix
\ref{app4} for a construction of (a representative of)
$[\eta_{\mathcal{D}}^{(n-p-1)}(z)]$. Now, let
\[
\eta_{\mathcal{D}}^{(n-p-1)}(z) := (\eta^{(0,n-p-1)}, \ldots ,
\eta^{(n-p-1,0)}, {\mathop \eta \limits^{\mathbb Z} {^{(n-p,-1)}}}
) \ ,
\]
be a representative of $[\eta_{\mathcal{D}}^{(n-p-1)}(z)]$ and
\[
\omega_{\mathcal{D}}^{(p)} := (\omega^{(0,p)}, \ldots ,
\omega^{(p,0)}, {\mathop \omega \limits^{\mathbb Z}{^{(p+1,-1)}}}
) \ ,
\]
a representative of a {\BD}  cohomology class
$[\omega_{\mathcal{D}}^{(p)}]$. Then a representative of the cup
product $[\omega_{\mathcal{D}}^{(p)}] \cup_{\cal{D}}
[\eta_{\mathcal{D}}^{(n-p-1)}(z)]$ is given by
\begin{eqnarray}\label{produitoe}
( \omega^{(0,p)}\ticup d\eta^{(0,n-p-1)}&,& \ldots ,
\omega^{(p,0)}\ticup d\eta^{(0,n-p-1)}, {\mathop \omega
\limits^{\mathbb Z}{^{(p+1,-1)}}}\ticup \eta^{(0,n-p-1)}, \ldots  \\
\nonumber  &,&  {\mathop \omega \limits^{\mathbb
Z}{^{(p+1,-1)}}}\ticup \eta^{(n-p-1,0)}, {\mathop \omega
\limits^{\mathbb Z}{^{(p+1,-1)}}}\ticup {\mathop \eta
\limits^{\mathbb Z}{^{(n-p,-1)}}}\, ).
\end{eqnarray}
The cup product ${\ticup}$ within the {\tch}-de Rahm complex is
defined in \cite{BT82}. In this {\BD} cohomology class, ${\mathop
\omega \limits^{\mathbb Z}{^{(p+1,-1)}}}\ticup {\mathop \eta
\limits^{\mathbb Z}{^{(n-p,-1)}}}$ is an integral {\tch}
$(n+1)$-cocycle which is necessarily trivial since the covering of
$M$ is simple. Hence
\begin{eqnarray}
{\mathop \omega \limits^{\mathbb Z}{^{(p+1,-1)}}}\ticup {\mathop
\eta \limits^{\mathbb Z}{^{(n-p,-1)}}} = \delta{\mathop \chi
\limits^{\mathbb Z}{^{(n,-1)}}}
\end{eqnarray}
for some integral {\tch} $n$-cochain ${\mathop\chi \limits^{\mathbb
Z}{^{(n,-1)}}}$. Accordingly, considering $M$ itself as a cycle we
can associate to it a Weil decomposition \footnote{Which is nothing
but a polyhedral decomposition of $M$, as defined in \cite{dR55}.}
\begin{eqnarray}{\label{Mdecomp}}
M^{\mathcal{W}} = (m_{(0,n)} , \ldots , m_{(k,n-k)} , \ldots ,
m_{(n,0)} ) ,
\end{eqnarray}
so that we obtain
\begin{eqnarray}\label{longdefform}
\int_{M} [\omega_{\mathcal{D}}^{(p)}] \cup_{\cal{D}}
[\eta_{\mathcal{D}}^{(n-p-1)}(z)] &\egzz& \sum_{k=0}^{p}
\int_{m_{(k,n-k)}}
\omega^{(k,p-k)}\ticup d\eta^{(0,n-p-1)} \\
\nonumber &+& \sum_{k=p+1}^{n} \int_{m_{(k,n-k)}} {\mathop \omega
\limits^{\mathbb Z}{^{(p+1,-1)}}}\ticup \eta^{(k-p-1,n-k)}
\\\nonumber &\egzz& \sum_{k=0}^{p} \int_{m_{(k,n-k)}}
\omega^{(k,p-k)}\ticup d\eta^{(0,n-p-1)} \ .
\end{eqnarray}

It has to be noted that not all representatives of
$[M^{\mathcal{W}}]$  and of $[\eta_{\mathcal{D}}^{(n-p-1)}(z)]$ are
suitable. Indeed, representatives of
$[\eta_{\mathcal{D}}^{(n-p-1)}(z)]$ are de Rham currents and so
cannot always be integrated on a singular chain. Strictly speaking,
the integration is possible only when the current and the chain are
transversal; this is the same problem as encountered in trying to
define the product of distributions. Intersection theory of chains
in $\mathbb{R}^n$ assures that there exist representatives of
$[M^{\mathcal{W}}]$ and $[z^{\mathcal{W}}_{(p)}]$ for which
(\ref{longdefform}) is well defined. More precisely the allowed
ambiguities on the representatives of the $m$'s and the $\eta$'s are
just those required to set the chains they represent in
{\em{``general position"}}, so that their intersection can be
defined (see for instance \cite{A77}). Then we can show that
(\ref{longdefform}) gives, up to integers, the same result as
(\ref{integclassomega}).

We shall refer to formula (\ref{longdefform}) as the ``\emph{Long
Formula}" which obviously allows to generalize the integration of
$[\omega_{\mathcal{D}}^{(p)}]$ over cycles in the sense that we
can now define the {\BD} product of $[\omega_{\mathcal{D}}^{(p)}]$
with any {\BD} cohomology class $[\eta_{\mathcal{D}}^{(n-p-1)}]$
(not necessarily representing a singular cycle) and integrate over
$M$. As an exercise, one can check that the two simple cases
presented in subsection (\ref{illus}) lead to the same results
when using the \emph{Long Formula}, instead of the {\emph
{Defining Formula}}.

\vspace{.5cm} \subsection{Smoothing}\label{smog}
\hspace{1.0cm}Instead of using singular chains as in the previous
construction we use here de Rham chains which are equivalence
classes of singular chains -\emph{for which the integrals of any
smooth form on $M$ are the same (}\cite{dR55}\emph{ p28)}-.
Accordingly we introduce de Rham integration currents $$T(z)_{\
k}^{n-p+k}$$ associated with $z_{(k,p-k)}$, elements of which can
be seen as $(n-p)$-forms with compact supports (and distributional
coefficients).
 In analogy with
(\ref{Wcycle}) we obtain a sequence of currents
\begin{eqnarray}\label{WTcycle}
 T_{\mathcal{W}}^{(p)}(z) =
({T(z)}_{\ 0}^{n-p} , \ldots , T(z)_{\ k}^{n-p+k} , \ldots ,
T(z)_{p}^{n} \ ) ,
\end{eqnarray}
 and the descent equations
\begin{eqnarray}\label{Wdescent0}
dT(z)_{\ k}^{n-p+k} = \partial T(z)_{\ k+1}^{n-p+k+1},
\end{eqnarray}
 for $ k =  1,...,p-1 $ and
\begin{eqnarray}\label{Wdescent2}
\partial T(z)_0  ^{n-p} = T(z)_{\ -1}^{n-p} := T(z)
\hspace{2cm} \int_{M} T(z)_p ^{n} \in [z_{(p,-1)}] \ ,
\end{eqnarray}
where $T(z)$ is the integration current of $z$ and $[z_{(p,-1)}]$
is the {\tch} homology class of $z$ in $M$. In terms of these de
Rham currents, the {\emph{Defining Formula}} reads
\begin{eqnarray}\label{integclassomega2}
\int_{z} [\omega_{\mathcal{D}}^{(p)}] \egzz \sum_{k=0}^{p}
\int_{M} T(z)_k^{n-p+k}\,\tiodot \, \omega^{(k,p-k)}
\end{eqnarray}
where we define: \begin{eqnarray} T(z)_k^{n-p+k}\, \tiodot \,
\omega^{(k,p-k)}=\frac{1}{(k+1)!}\, \sum_{\al_0, \cdots,
\al_k}\,T(z)^{n-p+k}_{k,\,{}_{ \al_0  \cdots \, \al_k}}\wedge
\omega^{(k,p-k)}_{{}_{\al_0  \cdots \, \al_k}}.
\end{eqnarray}

 \noindent As a special case, the whole cycle $M$
gives rise to a sequence
\begin{eqnarray}
T_{\mathcal{W}}(M) = (T(M)_0 ^0 , \ldots , T(M)_k ^k , \ldots ,
T(M)_n ^n ),
\end{eqnarray}
with
\begin{eqnarray}\label{Wdescent3}
dT(M)_{k}^{k} = \partial T(M)_{k+1}^{k+1}
\end{eqnarray}  for $ k =1,...,n-1 $ and
\begin{eqnarray}{\label{blab}}
\partial  T(M)_0 ^0 = T(M)_{-1} ^{\ 0}
 := T(M) = 1
\hspace{1cm};\  \int_{M} T(M)_n ^n \in [m_{(n,-1)}] \ ,
\end{eqnarray}
$[m_{(n,-1)}]$ being the {\tch} homology class of $M$.
Accordingly, the {\emph{Long Formula}} now reads
\begin{eqnarray}\label{longcurrent}
\int_{M} [\omega_{\mathcal{D}}^{(p)}] \cup_{\mathcal{D}}
[\eta_{\mathcal{D}}^{(n-p-1)}(z)] &\egzz& \sum_{k=0}^{p}\int_{M}
T(M)_k ^k \tiodot \lp\omega{^{(k,p-k)}}\ticup d
\eta^{(0,n-p-1)}\rp
 \\
\nonumber &+& \sum_{k=p+1}^{n} \int_{M}T(M)_k ^k \tiodot \lp
{\mathop \omega \limits^{\mathbb Z}{^{(p+1,-1)}}}\ticup
\eta^{(k-p-1,n-k)}\rp
 \ .
\end{eqnarray}

The allowed ambiguities of de Rham currents representing
$[T_{\mathcal{W}}(M)]$ are bigger than those implied by the Weil
descent in the decomposition of $M$,  except at the first and the
last steps -cf. ({\ref{blab}})-. Indeed, in (\ref{longcurrent}) an
ambiguity may be any de Rham current and not necessarily the
integration current of an integral chain as in the case of
(\ref{Mdecomp}), in particular it can be any $\mathbf{smooth}$
form (but still with compact support). This freedom on the
ambiguities allows us to smooth the $T(M)_k ^k$ currents occurring
in the {\emph{Long Formula}, replacing them by differential forms
induced by a partition of unity on $M$, as shown below.

Let us seek for sequences of (smooth) forms that satisfy the same
descent equations as $T_{\mathcal{W}}(M)$ and such that when
substituted  into (\ref{longcurrent}) they define the same
integrals. Concerning the descent equations, it is well-known (see
for instance \cite{AW52}) that a partition of unity on $M$,
subordinate to the simple covering ${\cal{U}}$ of $M$, gives rise to
a sequence of forms
\begin{eqnarray}\label{Thetadescent0}
\Theta_{\mathcal{W}}(M) := (\vartheta _0 ^0 , \ldots , \vartheta_k
^k , \ldots , \vartheta_n^n  \ ),
\end{eqnarray}
which satisfy homological descent equations
\begin{eqnarray}\label{Thetadescent1}
d\vartheta_k ^k  = \partial \vartheta_{k+1} ^{k+1}
\end{eqnarray}
$  k=1,...,n-1 $, as well as
\begin{eqnarray}\label{Thetadescent2}
\partial \vartheta_0 ^0 = \vartheta_{-1} ^{-1} = 1 \ .
\end{eqnarray}
Furthermore, since $M$ is supposed to be compact, the forms
$\vartheta_k ^k $ can all be chosen with compact supports in their
defining open sets. Due to the smoothness of all the components of
$\Theta_{\mathcal{W}} (M)$, the second constraint of (\ref{blab})
reads
\begin{eqnarray}\label{Wdescent4}
\int_{M} \vartheta_n ^n := t_{(n,-1)} + \partial r_{(n+1,-1)} \ ,
\end{eqnarray}
where $t_{(n,-1)}$ is an integral {\tch} cycle while
$r_{(n+1,-1)}$ is a \textbf{real} {\tch} chain. That is to say,
$\vartheta_n ^n$ defines an integral cycle up to a real boundary.
Using homological and cohomological descents, one can show that
$t_{(n,-1)} \in[m_{(n,-1)}]$. This is mainly due to the fact that
the integration of any closed $n$-form on $M$ can be performed by
means of a partition of unity on $M$.

Let us compare $T_{\mathcal{W}}(M)$ with $\Theta_{\mathcal{W}}(M)$
in order to replace $T_{\mathcal{W}}(M)$ by
$\Theta_{\mathcal{W}}(M)$ in (\ref{longcurrent}). To begin with,
\begin{eqnarray}\label{bla}
\partial \vartheta_0 ^0 - \partial T_0 ^0 = 0 \
\ \ \Rightarrow \ \ \vartheta_0 ^0 - T_0 ^0 = \partial R_1 ^0\ +
d_{-1} R_0 ^{-1},
\end{eqnarray}
with $\partial d_{-1} R_0 ^{-1} =0$,
 hence $\partial R_0 ^{-1}=0$.
As $M$ is connected $H_0 (M,\mathbb R) =0$, $R_0 ^{-1} =\partial
R_1 ^{-1}$. $T_0 ^0 $ can be replaced by $\vartheta_0 ^0$
 in (\ref{longcurrent}) since
\begin{equation}
\int_{M} d_{-1} R_0 ^{-1}\tiodot \lp \omega{^{(0,p-0)}}\ticup d
\eta^{(0,n-p-1)} \rp =\int_{M} d\, [  R_1 ^{-1} \tiodot \lp
\omega{^{(0,p-0)}}\ticup d \eta^{(0,n-p-1)} \rp ]=0.
\end{equation}
Thus $R_0 ^{-1}$ can be ignored in (\ref{bla}) and the first step
of the descent reads
\begin{eqnarray}
\partial (\vartheta_1 ^1 - T_1 ^1) =
d (\vartheta_0 ^0 -  T_0 ^0) = d \partial R_1 ^0 =
\partial d R_1 ^0 \ ,
\end{eqnarray}
so that
\begin{eqnarray}
\vartheta_1 ^1 - T_1 ^1 = d R_1 ^0 + \partial R_2 ^1 \ .
\end{eqnarray}
Similarly
\begin{eqnarray}
\vartheta_k ^k - T_k ^k = d R_k ^{k-1} + \partial R_{k+1} ^k \ ,\
\ k = {1, \ldots, n}.
\end{eqnarray}
Finally, the constraints (\ref{Wdescent2}) and (\ref{Wdescent4})
give
\begin{eqnarray}
\int_{M} (\vartheta_n ^n - T_n ^n) =  \partial \lambda_{n+1} ^{-1}
= \partial \int_{M} R_{n+1} ^n \ .
\end{eqnarray}

Now, if we replace $T_k ^k$ by $\vartheta_k ^k$ and its
ambiguities, the {\emph{Long Formula}} reads:
\begin{eqnarray} \hspace{-5 mm}
\int_{M} (...) &\egzz& \sum_{k=0}^{p} \int_{M} \vartheta_k
^k\tiodot \lp\omega^{(k,p-k)}\ticup d\eta^{(0,n-p-1)}\rp +
\sum_{k=p+1}^{n} \int_{M} \vartheta_k ^k\tiodot \lp {\mathop
\omega
\limits^{\mathbb Z}{^{(p+1,-1)}}}\ticup \eta^{(k-p-1,n-k)}\rp \nonumber \\
&+& \int_{M}
\partial R_{n+1} ^n  \tiodot\lp{\mathop \omega \limits^{\mathbb
Z}{^{(p+1,-1)}}}\ticup \eta^{(n-p-1,0)}\rp \ .
\end{eqnarray}
The last term in this equation gives
\begin{eqnarray}\hspace{-7 mm}
&&\int_{M} R_{n+1} ^n  \tiodot\delta\, ({\mathop \omega
\limits^{\mathbb Z}{^{(p+1,-1)}}}\ticup \eta^{(n-p-1,0)}) =
\int_{M} R_{n+1} ^n\tiodot \delta {\mathop \chi \limits^{\mathbb
Z}{^{(n,-1)}}} \nonumber
\\ &=& \int_{M}
\partial R_{n+1} ^n\tiodot  {\mathop \chi \limits^{\mathbb
Z}{^{(n,-1)}}}  = \int_{M} (\vartheta_n ^n - T_n ^n)\tiodot
{\mathop \chi \limits^{\mathbb Z}{^{(n,-1)}}}  \ .
\end{eqnarray}
Since all integrals of  $T_n ^n$'s are integers, we obtain
\begin{eqnarray}
\int_{M} R_{n+1} ^n \tiodot \delta\, ({\mathop \omega
\limits^{\mathbb Z}{^{(p+1,-1)}}}\ticup \eta^{(n-p-1,0)}) \egzz
\int_{M} \vartheta_n ^n\tiodot {\mathop \chi \limits^{\mathbb
Z}{^{(n,-1)}}} \nonumber \ ,
\end{eqnarray}
so that the {\emph{(smoothed) Long Formula}} reads
\begin{eqnarray}{\label{smoothlong}}
\hspace{-5 mm}\int_{z} [\omega_{\mathcal{D}}^{(p)}]  &\egzz&
\sum_{k=0}^{p} \int_{M} \vartheta_k ^k \tiodot
\lp\omega^{(k,p-k)}\ticup d\eta^{(0,n-p-1)}\rp
\\ &+& \sum_{k=p+1}^{n} \int_{M} \vartheta_k ^k \tiodot \lp
{\mathop \omega \limits^{\mathbb Z}{^{(p+1,-1)}}}\ticup
\eta^{(k-p-1,n-k)}\rp +  \int_{M} \vartheta_n ^n\tiodot  {\mathop
\chi \limits^{\mathbb Z}{^{(n,-1)}}} \nonumber \ .
\end{eqnarray}

\vspace{.5cm} Let us make some final remarks. First, if the simple
covering ${\cal{U}}$ of $M$ is such that all intersections of
order larger than $n+1$ are empty - we shall say that $\cal{U}$ is
\emph{``excellent"}- we deduce that
\begin{equation}
\int_{M} \vartheta_n ^n\tiodot  {\mathop \chi \limits^{\mathbb
Z}{^{(n,-1)}}}  \in \mathbb{Z}, \nonumber
\end{equation}
which leads to
\begin{eqnarray}
\int_{z} [\omega_{\mathcal{D}}^{(p)}]  &\egzz& \sum_{k=0}^{p}
\int_{M} \vartheta_k ^k \tiodot \lp\omega^{(k,p-k)}\ticup
d\eta^{(0,n-p-1)}\rp
\\ &+& \sum_{k=p+1}^{n} \int_{M} \vartheta_k ^k \tiodot \lp
{\mathop \omega \limits^{\mathbb Z}{^{(p+1,-1)}}}\ticup
\eta^{(k-p-1,n-k)} \rp.
\end{eqnarray}

In other words, with respect to an excellent covering of $M$, the
$\vartheta_k ^k$'s play the role of the integration currents
$T(M)_k ^k$ of the Weil descent of $M$.

\vspace{.5cm}

Second, the previous construction, {\emph{i.e.}} the smoothing,
cannot be applied to the {\em{Defining Formula}} without care.
Indeed, a simple covering of $M$ does not always induce a simple
covering on the $p$-cycle $z_p$, so that, although a
$(p+1)$-cocycle on $M$ reduces to a $(p+1)$-cocycle on $z_p$, this
cocycle is not necessarily trivial. Therefore we cannot establish
a smoothed {\em{Defining Formula}} in full generality. However,
let us assume that $z_p$ admits a tubular neighborhood,
$\mathcal{V}_z$, such that ${\cal{U}}_{\mid\mathcal{V}_z}$ - the
restriction to $\mathcal{V}_z$ of the simple covering ${\cal{U}}$
of $M$ - is also simple. Then, as a tubular neighborhood,
$\mathcal{V}_z$ has necessarily the same cohomology as $z_p$, and
since ${\cal{U}}_{\mid\mathcal{V}_z}$ is simple, this cohomology
is also the \tch\ cohomology of ${\cal{U}}_{\mid\mathcal{V}_z}$.
In particular the \tch\ $(p+1)$-cocycle, ${\mathop \omega
\limits^{\mathbb Z}{^{(p+1,-1)}}}$, on $M$ is also a
$(p+1)$-cocycle on $\mathcal{V}_z$ and  is necessarily trivial on
it, that is: ${\mathop \omega \limits^{\mathbb Z}{^{(p+1,-1)}}} =
\delta \teger{\varpi} {}^{(p,-1)}$ for some integral \tch\
$p$-cochain $\teger{\varpi} {}^{(p,-1)}$, just as in the case of
the {\em{Long Formula}}. With all this, a natural candidate for a
smoothed {\em{ Defining Formula}} would be
\[\int_{z_p} [\omega_{\mathcal{D}}^{(p)}] \egzz \sum_{k=0}^{p-1} \int_{z_p}
\vartheta_k^{k}\tiodot \omega^{(k,p-k)} + \int_{z_p}
\vartheta_p^{p}\tiodot (\omega^{(p,0)} - \teger{\varpi}
{}^{(p,-1)}) \ , \] which compares to the smoothed {\em{Long
Formula}} (\ref{smoothlong}).

\vspace{.5cm}

As a third remark, one can wonder what is the relation between the
{\em{Defining Formulas}} and the decomposition $A = A_0 + \al$
used in section 2 in the case of $U(1)$-connections. Let us
consider two \BD\ classes, $[\omega_\mathcal{D}]$ and
$[\chi_\mathcal{D}]$, representing the same \tch\ cohomology
class, $[\check{\xi}]$, as detailed in appendix A.6. We know that
$[\omega_\mathcal{D}]$ and $[\chi_\mathcal{D}]$ differ by a \BD\
class, $[(\delta {\al})_\mathcal{D}]$ coming from a global form
$\al$ on $M$.
 This exactly corresponds to the standard decomposition $A =
A_0 + \al$ for $U(1)$-connections met in section 2. This can also
be seen at the level of the integrals : choose representatives
$(\omega^{(0,p)}, \cdots , \omega^{(p,0)}, \teger{\omega}
{}^{(p+1,-1)})$ and $(\chi^{(0,p)}, \cdots , \chi^{(p,0)},
\teger{\chi} {}^{(p+1,-1)})$ of $[\omega_\mathcal{D}]$ and
$[\chi_\mathcal{D}]$ respectively, and write the previous
decomposition
\begin{eqnarray}
(\chi^{(0,p)}, \cdots , \chi^{(p,0)}, \teger{\chi} {}^{(p+1,-1)})
&=& (\omega^{(0,p)}, \cdots , \omega^{(p,0)}, \teger{\omega}
{}^{(p+1,-1)}) + (0, \cdots , 0, \delta \al) \nonumber\\ &&+ \
D(q^{(0,p-1)}, \cdots , q^{(p-1,0)}, \teger{q} {}^{(p,-1)}),
\end{eqnarray}
for some $[q_\mathcal{D}]$. By assumption $\teger{\chi}
{}^{(p+1,-1)}$ and $\teger{\omega} {}^{(p+1,-1)}$ are
cohomologous, so
\[\int_{z_p} [\chi_\mathcal{D}] \egzz \int_{z_p} [\omega_\mathcal{D}] + \int_{z_p} \al. \]
This result also means that the standard decomposition $A = A_0 +
\al$ of $U(1)$-connections, extends to any generalized
$p$-connection.

A final remark on notations, we could have denoted the integral
over $z_p$ of the class $[\omega_{\mathcal{D}}^{(p)}]$ simply as:
\begin{eqnarray} \label{longo}
\int_{z_p} [\omega_{\mathcal{D}}^{(p)}]  &\egzz&
<[\omega_{\mathcal{D}}^{(p)}],[z^{\mathcal{W}}_{(p)}]>
 \\ &\egzz& <\omega^{(0,p)} + \cdots
\omega^{(p,0)} + {\mathop \omega \limits^{\mathbb Z}{^{(p+1,-1)}}}
, z_{(0,p)} + \cdots  + z_{(p,0)}> \nonumber
\\&\egzz& <[\omega_{\mathcal{D}}^{(p)}] \cup_{\cal{D}}
[\eta_{\mathcal{D}}^{(n-p-1)}(z)], \ M> \nonumber
\end{eqnarray}
which has the advantage to make easier the proof of independence
with respect to the various representatives.

\sect{Integration of {\BD} classes with distributional
coefficients}
\hspace{1.0cm} In any quantization procedure,
$\omega$ will be by nature distributional and integration over a
cycle will, in general, be ill defined so that the integration
current of the cycle will have to be replaced by some regularized
form. This is the situation which has been exhibited in the
example of section 2. A canonical way to perform such an operation
for $[\omega_{\mathcal{D}}^{(p)}]$ of distributional character is
to use formula (\ref{smoothlong}, \ref{longo}) with
$z^{\mathcal{W}}_{(p)}$ replaced by a smooth {\BD} class
$[j_{\mathcal{D}}^{(n-p-1)}]$, the integration formula being
\begin{eqnarray}\label{jint}
 <[\omega_{\mathcal{D}}^{(p)}] \cup_{\cal{D}}
[j_{\mathcal{D}}^{(n-p-1)}], \ M> &\egzz& \sum_{k=0}^{p} \int_{M}
\vartheta_k ^k \tiodot \lp\omega^{(k,p-k)}\ticup dj^{(0,n-p-1)}\rp
\\ &+& \sum_{k=p+1}^{n} \int_{M} \vartheta_k ^k \tiodot \lp
{\mathop \omega \limits^{\mathbb Z}{^{(p+1,-1)}}}\ticup
j^{(k-p-1,n-k)}\rp +  \int_{M} \vartheta_n ^n\tiodot  {\mathop
\chi \limits^{\mathbb Z}{^{(n,-1)}}} \nonumber \ . \end{eqnarray}
Note that (\ref{jint}) is - $\mathbf{mod\ \mathbb{Z}}$ ! -
symmetric in $[\omega_{\mathcal{D}}^{(p)}]$ and
$[j_{\mathcal{D}}^{(n-p-1)}]$, as can be easily verified. Whereas
we have shown that to the current of a cycle $z_p$ is associated a
special {\BD} class $[\eta_{\mathcal{D}}^{(n-p-1)}(z)]$, the map
$z_p \rightarrow [\eta_{\mathcal{D}}^{(n-p-1)}(z)]$ being
analogous to the cycle map in \cite{B85}, we do not know of such
an assignment in the case of a smoothed version. It is expected
that after renormalization some of the characteristics of the
regularized class $[j_{\mathcal{D}}^{(n-p-1)}]$ will survive.

\section{Conclusions}
\hspace{1.0cm}We have described in some details a class of
topological actions which are ``topological" in the sense that
they are defined modulo ``integers", a situation repeatedly met in
semi classical treatments of various field theories involving
particular geometries (mostly gauge theories, including gravity).
They are described by integral formulae which involve refinements
of closed differential forms with integral periods named {\BD}
cohomology classes. The integrals are written as pairings of two
such classes in such a way that one of them may have a
distributional character as demanded in most field theory
contexts.

\newpage
\appendix

\sect{\BD\ cohomology}
\label{appDB}
\hspace{1.0cm} We have not been
able to find an elementary discussion of \BD\ cohomology in the
mathematical literature. The purpose of this appendix is to fill
in this gap, concentrating on the computation of \BD\ cohomology
and on the proof of its independence upon the covering. For more
algebraic expos\'es we refer to \cite{EV88, Bry93}.

\subsection{Definitions and notations}
\hspace{1.0cm}As in the main text, $M$ denotes a compact differentiable manifold
of dimension $n$, and $\{\mathcal{U}_{\alpha}\}_{\alpha \in I}$ a
simple covering\footnote{Such an open covering is alternatively
called a {\em
  good} covering in \cite{BT82}. This means that any finite intersection
of $\mathcal{U}_{\alpha}$'s, $\mathcal{U}_{{\alpha_0} \cdots
  {\alpha_q}}=\mathcal{U}_{\alpha_0}\cap \cdots \cap
\mathcal{U}_{\alpha_q}  \ , \ (\alpha_0,\cdots,\alpha_q)\in
I^{q+1},$   is either empty or diffeomorphic to $\mathbb{R}^n$. }
of $M,\ M={\cup_{\alpha \in I}\mathcal{U}_{\alpha}}$. A \tch\
cochain of degree $k$ with values in an abelian group $G$ is a
collection of elements $c_{\alpha_0\cdots\alpha_k}$ of $G$, one
for each intersection $\mathcal{U}_{\alpha_0\cdots\alpha_k}$,
which is totally antisymmetric in all its indices and vanishes on
empty intersections. A \tch\ cochain of degree $-1$ is a constant
map from $M$ to $G$.

The \tch\ differential, $\delta$, maps $(k-1)$-cochains to
$k$-cochains and squares to $0$. Acting on $(-1)$-cochains,
$\delta$ is the restriction : $(\delta c)_{\alpha_0}=c\ $ on any
non empty $\mathcal{U}_{\alpha_0}$. For $k \geq 1$, if
$c_{\alpha_0\cdots\alpha_{k-1}}$ is a $(k-1)$-cochain and
$\mathcal{U}_{\alpha_0\cdots\alpha_k} \neq \emptyset$,
\begin{equation} \label{eq:cechdiff}
(\delta c)_{\alpha_0\cdots\alpha_{k}} =\sum_{i=0} ^{k}\,
(-)^{i}c_{\al_0\cdots\widehat{\alpha}_i\cdots \alpha_k}
\end{equation}
were the $\widehat{\hspace{.2cm}}$ means omission.

The elements in the kernel of $\delta$ are \tch\ cocycles, those
in the image of $\delta$ are  \tch\ coboundaries.

In the sequel, we shall have no use of general abelian groups $G$,
but $\mathbb{R}$ ( for real \tch\ cochains), $\mathbb{Z}$ ( for
integral \tch\ cochains) and $\mathbb{R}/\mathbb{Z}$ will play
preferred roles.

One can  also consider  \tch\ cochains where each
$c_{\alpha_0\cdots\alpha_k}$ is a differential $l$-form defined on
$\mathcal{U}_{\alpha_0\cdots\alpha_k}$; such cochains are often
referred to as \tch -de Rham cochains of bidegree $(k,l)$. In
\tch\ degree $-1$, we retrieve global differential $l$-forms
defined on $M$ and $\delta$ is still defined by restriction. On
these ``extended" {\tch}
 $(k-1)$-cochains, $k \geq 1$, the action of $\delta$ is still
given by (\ref{eq:cechdiff}) except for an overall multiplicative
factor $(-)^{l+1}$ on the right hand side : each term makes sense
with the proviso that it is restricted to the corresponding
$(k+1)$-fold intersection . This leads to the space\footnote{A
more appropriate language for this
  setting involves sheaves, but we shall not use the corresponding terminology.}  denoted
by $\check{C}^{(k)}({\mathcal{U}},\Omega^l(M))$ in the main text.
To save space in this appendix, we shall denote it simply by $\Om
^{(k,l)}(\mathbb{R})$, because most of the time $M$ and
$\mathcal{U}$ will be fixed.

By convention, a ``purely \tch\ " cochain with constant
coefficients (in a subgroup $G$ of $\mathbb{R}$) receives form
degree $-1$, so it belongs to $\Om ^{(k,-1)}(G)$. The de Rham
differential $d$ maps $\Om ^{(k,l)}(G)$ into $\Om ^{(k,l+1)}(G)$
for $k \geq 0\ ${\footnote{The sign factor $(-1)^{l+1}$ insures
that $d\, \delta +\delta\, d=0$.}}. We extend $d$ to $(-1)$-forms
as the injection which maps an element of $G \subset \mathbb{R}$
to the corresponding constant function. This is sometimes denoted
by the symbol $d_{-1}$. This extension still satisfies $d ^{\,
2}=0$ .

Later in the appendix, we shall need to compare several simple
coverings. Suppose that the simple covering $\mathcal{V}=
\{\mathcal{V}_{\sigma}\}_{\sigma \in J}$ of $M$ is a refinement of
the simple covering $\mathcal{U}= \{\mathcal{U}_{\alpha}\}_{\alpha
\in I}$ : this means that there is the restriction map $r : J
\longrightarrow I$ such that $\mathcal{V}_{\sigma} \subset
\mathcal{U}_{r(\sigma)}$ for all indices $\sigma \in J$. A \tch\
$k$-cochain, $c$, for $\mathcal{U}$ can be restricted to
$\mathcal{V}$ : if the intersection $\mathcal{V}_{\sigma_0 \cdots
\sigma_k}$ is nonempty, then so is $\mathcal{U}_{r(\sigma_0)
\cdots r(\sigma_k)}$, and
$$r(c)_{\sigma_0\cdots\sigma_{k}}\equiv \left({c_{r(\sigma_0)
      \cdots r(\sigma_k)}}\right)_{|\mathcal{V}_{\sigma_0 \cdots
    \sigma_k}}.$$
The \tch\ and de Rham differentials commute with restriction,
{\emph{i.e.}} $\delta\circ r=r\circ \delta$ (it being understood
that the \tch\ differential on the left-hand side is for the
covering $\mathcal{V}$ and on the right-hand side for the covering
$\mathcal{U}$) and $d\circ r= r\circ d$.

\subsection{\BD\  cochains}
\hspace{1.0cm}Take an integer $0 \leq p \leq n+1$ ( $n$ the dimension of the
manifold) and consider the double complex
$$ \xymatrix{
\Om ^{(0,-1)}(\mathbb{Z})  \ar[r]^{d_{-1}} \ar[d]^-\delta &\Om
^{(0,0)}(\mathbb{R}) \ar[r]^-d \ar[d]^-\delta &\cdots \ar[r]^-d
&\Om ^{(0,p-1)}(\mathbb{R})
\ar[r]^-0 \ar[d]^-\delta &0 \\
\Om ^{(1,-1)}(\mathbb{Z})  \ar[r]^{d_{-1}} \ar[d]^-\delta &\Om
^{(1,0)}(\mathbb{R}) \ar[r]^-d \ar[d]^-\delta &\cdots  \ar[r]^-d
&\Om ^{(1,p-1)}(\mathbb{R}) \ar[d]^-\delta \ar[r]^-0  &0\\  \Om
^{(2,-1)}(\mathbb{Z}) \ar[d]^-\delta \ar[r]^{d_{-1}} &\Om
^{(2,0)}(\mathbb{R}) \ar[d]^-\delta \ar[r]^-d &\cdots \ar[r]^-d
&\Om ^{(2,p-1)}(\mathbb{R}) \ar[r]^-0 \ar[d]^-\delta &0 \\\vdots &
\vdots  & &\vdots}
$$
The columns of this diagram form  standard \tch\ complexes. The
rows are Deligne complexes of index $p$, that is de Rham complexes
extended to the left by $d_{-1}$ (the injection of integral
constants into real functions) and truncated on the right at
$(p-1)$-forms by the 0 map. We denote by $\td$ this modified
differential, to avoid confusion with the de Rham differential,
$d$.

We  build a new ``diagonal complex'' from this double complex. The
space $C_p^q$ of \BD\ cochains of degree $q \geq 0$ (with fixed
index $p$) is defined by
$$
C_p^{q}= \left\{
\begin{array}{lllllc}
\Om ^{(q,-1)}(\mathbb{Z})+ {\sum\limits_{k=1}^{q}} \Om
^{(q-k,k-1)}
(\mathbb{R})& \mbox{for } 0 \leq q < p \\  \\
\Om ^{(p,-1)}(\mathbb{Z})+ {\sum\limits_{k=1}^{p}} \Om
^{(p-k,k-1)}
(\mathbb{R})& \mbox{for } \ q = p \\  \\
 \Om^{(q,-1)}(\mathbb{Z})+\sum\limits_{k=1} ^p \Om ^{(q-k,k-1)} (\mathbb{R})
&  \mbox{for } q > p \end{array} \right. .$$ Elements of these
spaces are respectively represented by the following sequences :
\[c=(c^{(0,q-1)},\cdots,\cg {}^{(q,-1)}),c=(c^{(0,p-1)},\cdots,\cg
{}^{(p,-1)}),
\\c=(c^{(q-p,p-1)},\cdots,\cg {}^{(q,-1)}),\]with the last
element $\mathbb Z$-valued \footnote{Our complex contains $\Om
^{(q,-1)}(\mathbb{Z})$ while in the literature one usually finds
$\Om ^{(q,-1)}(\mathbb{Z}(p))$, where $\mathbb{Z}(p) =
(2i\pi)\mathbb{Z}$. This difference is irrelevant for our
purpose.}.

 We set $\mathcal{C}_p=C_p^0\oplus C_p^1\oplus \cdots$.

The operator $D=\td+\delta$ maps $C_p^q$ to $C_p^{q+1}$ and, due
to the sign convention in the definition of $\delta$ on $l$-forms,
$D^2=0$. The complex $(\mathcal{C}_p,D)$ is called the \BD\
complex\footnote{A
  better notation would be $(\mathcal{C}_p(M),\mathcal{U},D)$.}, and
the elements of $\mathcal{C}_p$ \BD\ cochains. We write
$Z_p^q=\left\{\mbox{Ker } D : C_{p}^{q}\rightarrow
  C_{p}^{q+1}\right\}$ (resp. $B_p^q=\left\{\mbox{Im } D :
  C_{p}^{q-1}\rightarrow C_{p}^{q}\right\}$) for the space of \BD\
cocycles (resp. coboundaries).

We are interested in the cohomology of $(\mathcal{C}_p,D)$. A
priori, it depends on the covering, but we shall see later that
the cohomologies for simple coverings are canonically isomorphic.

The projection  $\pi : C^q_p \rightarrow \Om
^{(q,-1)}(\mathbb{Z})$ gives a chain map

$$ \xymatrix{\cdots \ar[r]^D &C^q_p  \ar[r]^D \ar[d]^\pi &C^{q+1}_p
\ar[r]^D \ar[d]^\pi &\cdots \\
\cdots \ar[r]^-\delta &\Om ^{(q,-1)}(\mathbb{Z})  \ar[r]^\delta
&\Om ^{(q+1,-1)}(\mathbb{Z}) \ar[r]^-{\delta} &\cdots} $$

 so that in all cases, there is a canonical map
$H^{q}(\mathcal{C}_p,D) \rightarrow
H^{q}_{\check{C}ech}(M,\mathbb{Z})$. The computation of
$H^{q}(\mathcal{C}_p,D)$ goes along different lines whether $q
\leq p-1$ or $q>p-1$.
\subsection{Computation of $H^{q}(\mathcal{C}_p,D)$, $q <
p$}
\hspace{1.0cm}In this case, we use the Poincar\'e lemma for
differential forms (ensuring that for forms of nonnegative degree,
the de Rham cohomology is locally trivial) to show that

\[ H^{q}(\mathcal{C}_p,D) \ \ \simeq\ \  H^{q-1}_{\check{C}ech}(M,\mathbb{R}
/\mathbb{Z}) \ \ \ \ \ \  (I)\] (the isomorphism is canonical). In
particular, the canonical map
\[ H^{q}(\mathcal{C}_p,D) \rightarrow H^{q}_{\check{C}ech}(M,\mathbb{Z})\]
maps $H^{q}(\mathcal{C}_p,D)$ onto the subgroup
$H^{q}_{\check{C}ech}(M, \mathbb{Z})_{torsion}$ of torsion
classes.

{\textbf{Proof:}} Suppose
$c=(c^{(0,q-1)},c^{(1,q-2)},\cdots,c^{(q-1,0)},\cg {}^{(q,-1)})$
is a \BD\ cocycle. This implies that $\td c^{(0,q-1)}=0$, and
since $q \leq p-1$, the operator $\td$ in this equation is the
standard de Rham differential.  So, by the Poincar\'e lemma, there
is an element $\rho ^{(0,q-2)} \in \Om ^{(0,q-2)}(\mathbb{R})$
such that $c^{(0,q-1)}+\td \rho ^{(0,q-2)}=0$. Accordingly the
cocycle $c$ is cohomologous to the cocycle $$c+D\rho
^{(0,q-2)}=(0,\underline{c}^{(1,q-2)}, \cdots,c^{(q-1,0)},\cg
{}^{(q,-1)}),$$ where $\underline{c}^{(1,q-2)} \equiv c^{(1,q-2)}+
\delta \rho ^{(0,q-2)} $.

 The cocycle condition for $c+D\rho ^{(0,q-2)}$
yields $d\underline{c}^{(1,q-2)}=0$ were $d$ is the standard
exterior derivative. The procedure can be iterated to show that
the cohomology class of $c$ contains a representative of the form
$$(0,\cdots,0,\underline{c}^{(q-1,0)},\cg {}^{(q,-1)})$$
 with the standard descent equations fulfilled :
$$
d\underline{c}^{(q-1,0)}=0,\ \ \delta\underline{c}^{(q-1,0)}=
d_{-1}\cg {}^{(q,-1)}, \ \ \delta \cg {}^{(q,-1)}=0.
$$
The first equation just tells that
 $\underline{c}^{(q-1,0)} = d_{-1} \rho ^{(q-1,-1)}$, where the components
  $\rho ^{(q-1,-1)}$ are real constants. This, combined with the second
  equation,
implies that the integral \tch\ cocycle $\cg {}^{(q,-1)}$ is exact
as a real cocycle, so that it represents a torsion class.

Reduction modulo $1$ turns $\rho^{(q-1,-1)}$ into an $\mathbb{R}
/\mathbb{Z}$ \tch\ cocycle and the ambiguity on
$\underline{c}^{(q-1,0)} (\mbox{mod }1)$ is a \tch\ coboundary. So
we have proved the announced result, $(I)$, which is also the
content of the following exact sequence \cite{Bry93}
$$\xymatrix@1{
0 \ar[r] &H^{q-1}(M, {\mathbb Z}(p)) \ar[r] &H^{q-1}(M, \mathbb
R)\ar[r] &H^q ({\mathcal C}_p, D) \ar[r] & H^q(M, {\mathbb
Z}(p))_{torsion}\ar[r] &0.}$$

\subsection{The \tch\ homotopy operator}
\hspace{1.0cm}Here we introduce the \tch\ homotopy operator that
we shall need
 to compute $H^{q}(\mathcal{C}_p,D)$ in the special cases $q\geq p$.
This  homotopy\footnote{Ensuring that the \tch\ cohomology for
forms of nonnegative degree is trivial.} operator, which depends
on a partition of unity defined on $M$, is instrumental to
establish the generalized Mayer-Vietoris exact sequence, the
\tch-de Rham isomorphism and the \emph{Collating Formula}
\cite{BT82}, a construction we illustrate below.

\subsubsection{The $K$ operator on the enlarged double complex}

Consider the following double complex :

$$
\xymatrix{
 &  \Om ^{(-1,0)}(\mathbb{R})
 \ar[r]^-d \ar[d]^\delta &\cdots  \ar[r]^-d
&\Om^{(-1,p-1)}(\mathbb{R}) \ar[r]^-0 \ar[d]^\delta &0\\
\Om ^{(0,-1)}(\mathbb{Z}) \ar[r]^{d_{-1}} \ar[d]^\delta &\Om
^{(0,0)}(\mathbb{R})  \ar[r]^-d \ar[d]^\delta &\cdots  \ar[r]^-d
&\Om ^{(0,p-1)}(\mathbb{R})
\ar[r]^-0 \ar[d]^\delta &0 \\
\Om ^{(1,-1)}(\mathbb{Z}) \ar[r]^{d_{-1}} \ar[d]^\delta &\Om
^{(1,0)}(\mathbb{R}) \ar[r]^-d \ar[d]^\delta &\cdots  \ar[r]^-d &
\Om ^{(1,p-1)}(\mathbb{R})
\ar[r]^-0 \ar[d]^\delta &0 \\
\Om ^{(2,-1)}(\mathbb{Z}) \ar[d]^\delta \ar[r]^{d_{-1}}  &\Om
^{(2,0)}(\mathbb{R})\ar[d]^\delta \ar[r]^-d &\cdots  \ar[r]^-d
&\Om ^{(2,p-1)}(\mathbb{R})\ar[d]^\delta
\ar[r]^-0 &0 \\
\vdots &  \vdots & &  \vdots}
$$
where the de Rham  complex of global differential forms truncated
at degree $(p-1)$ has been added at the top. We extend the
definition of $D$ to this enlarged complex.

Let us choose a partition of unity $\th _{\alpha}$ subordinate to
the simple covering $\{\mathcal{U}_{\alpha}\}_{\alpha \in I}$ of
$M$: each $\th _{\alpha}$ is a (smooth) non-negative function on
$M$ with compact support in $\mathcal{U}_{\alpha}$, and
$\sum_{\alpha} \th_{\alpha}$ is the constant function $1$ on $M$.
On the enlarged complex, define an operator $K$ (depending on the
chosen partition of unity) as follows.

Take $c =\{c_{\alpha_0\cdots\alpha_{k}}\}\in \Om
^{(k,l)}(\mathbb{R})$, $k,l \geq 0$. Due to the support properties
of the $\th_{\alpha}$'s, $c_{\alpha_0\cdots\alpha_k} \cdot \th
_{\alpha_k}$ (extended by $0$ outside $\mathcal{U}_{\alpha_k}$) is
a smooth differential form in each nonempty
$\mathcal{U}_{\alpha_0\cdots \alpha_{k-1}}$. Let $Kc \equiv
\{(-)^{l+1}\sum_{\alpha_k} c_{\alpha_0\cdots\alpha_k} \cdot \th
  _{\alpha_k}\} \in \Om ^{(k-1,l)}(\mathbb{R})$.

For $c \in \Om ^{(-1,l)}(\mathbb{R})$, $l \geq 0$, set $Kc \equiv
0$, and for $c \in \Om ^{(k,-1)}(\mathbb{Z})$, $k \geq 0$, set $Kc
\equiv Kd_{-1}c \in \Om ^{(k-1,0)}(\mathbb{R})$.

Though we shall not try to compute its homology, note that $K^2=0$
so $K$ is a boundary operator (or equivalently a co-differential).

\subsubsection{The homotopy property and the fundamental identity}

Algebraic manipulations show that $K \delta+\delta K$ is the
identity operator on $\Om ^{(k,l)}(\mathbb{R})$, $k \geq -1, l\geq
0$ and $d_{-1}$ on $\Om ^{(k,-1)}(\mathbb{Z})$, $k \geq 0$. In
particular, in the enlarged double complex, the vertical \tch\
complexes in nonnegative de Rham degree have vanishing \tch\
cohomology, since $K$ is a homotopy operator.

Acting on the enlarged double complex, $K\td$ lowers the \tch\
degree by one unit, so $K\td$ is locally nilpotent and $1+K\td$ is
invertible : locally the geometric series for $(1+K\td)^{-1}$
stops after a finite number of terms. Moreover, as a consequence
of
\[(1+K\td)(\td+\delta)-\delta(1+K\td) = \td+K\td\delta-\delta K\td
= (1-K\delta-\delta K)\td=0,\] (the first equality uses
$\td^{\,2}=0$, the second $\td\delta=-\delta\td$ and the last one
that the image of $\td$ lives in de Rham degree $\geq 0$ where
$K\delta+\delta K=1$) one derives that on the enlarged double
complex, $D$ and $\delta$ are conjugate, that is
\[(1+K\td)D=\delta(1+K\td). \ \ \ (\heartsuit) \]
   This fundamental identity $(\heartsuit)$ is at the
heart of the computation of the \BD\ cohomology when $q \geq p$ as
shown later in \ref{hpplus} and {\ref{hpegal}. It can also be
useful in other contexts as illustrated below.

\subsubsection{Relation with the \tch -de Rham isomorphism}
\label{cechderahmisom}
Suppose that in the first column of the
enlarged complex we replace the coefficient group $\mathbb{Z}$ by
$\mathbb{R}$, and that we take $p=n+1$, $n$ the dimension of the
manifold, so that the lines are usual de Rham complexes, hence
$\td =d$ in this enlarged context and the ($\heartsuit$) identity
can be written $(1+Kd)D=\delta(1+Kd)$. This double complex is a
{\tch}-de Rham complex with differential $D=d+\delta$ and of
course $q<p=n+1 \, $. In the sequel this is the complex we have in
mind when we refer to {\tch}-de Rham cochains, cocycles or
coboundaries.

On the one hand if $c^{(q,-1)}\in \Om ^{(q,-1)}(\mathbb{R})$ is a
\tch\ cocycle, it is a $D$-cocycle, hence its top component
$(-Kd)^{q+1}c^{(q,-1)}$ is a global closed $q$-form, {\emph{i.e.}}
a de Rham $q$-cocycle.

On the other hand if $c^{(q,-1)}$ is a \tch\ coboundary,
$c^{(q,-1)}=\delta \gamma^{(q-1,-1)}$ for some $\gamma^{(q-1,-1)}
\in \Om ^{(q-1,-1)}(\mathbb{R})$, then using ($\heartsuit$)
 $(1+Kd)^{-1}c^{(q,-1)}=D(1+Kd)^{-1} \gamma^{(q-1,-1)}$ is a
$D$-coboundary. Identifying top  form components,
$(-Kd)^{q+1}c^{(q,-1)}$ is a de Rham coboundary $d(-Kd)^{q}
\,\gamma^{(q-1,-1)}$.

Finally, if $c= (c^{(-1,q)}, \cdots, c^{(q-1,0)},c^{(q,-1)})$ is a
$D$-cocycle, $c^{(q,-1)}$ is a \tch\ cocycle, $c^{(-1,q)}$ is a
closed global de Rham $q$-form, and $c$ is $D$-cohomologous to
$(1+Kd)^{-1}c^{(q,-1)}$. Indeed, start from $D(1+Kd)^{-1}K
(c-c^{(q,-1)})=(1+Kd)^{-1}\delta K (c-c^{(q,-1)})$, a consequence
of the ($\heartsuit$) identity. As $c-c^{(q,-1)}$ has no component
in de Rham degree $-1$, $\delta K (c-c^{(q,-1)})=(1-
K\delta)(c-c^{(q,-1)})$ by the homotopy property.  By $Dc=0=\delta
c^{(q,-1)}$, we obtain finally that $\delta K
(c-c^{(q,-1)})=(1+Kd)c-c^{(q,-1)}$. Multiplication by
$(1+Kd)^{-1}$ leads to
 \[c=(1+Kd)^{-1}c^{(q,-1)} +D(1+Kd)^{-1}K
(c-c^{(q,-1)}), \ \ (*) \] proving that $c$ is $D$-cohomologous to
$(1+Kd)^{-1}c^{(q,-1)}$. This implies that $c^{(-1,q)}$ is
$d$-cohomologous to $(-Kd)^{q+1}c^{(q,-1)}$, explicitly,
$$c^{(-1,q)}=(-Kd)^{q+1}c^{(q,-1)}+d\left(K
  \sum_{r=0}^{q-1}(-dK)^rc^{(r,q-1-r)}\right),$$
which is the famous \emph{Collating Formula}; see e.g.
\cite{BT82}, where it is used to prove that $c^{(q,-1)}\rightarrow
(-Kd)^{q+1}c^{(q,-1)}$ which maps (real) \tch\ cocycles to de Rham
cocycles and (real) \tch\ coboundaries to de Rham coboundaries
induces an isomorphism in cohomology. With notations closer to the
ones used in the main text, the \emph{Collating Formula} can be
rewritten {\footnote{ cf.
(\ref{Thetadescent0})-(\ref{Thetadescent2}) in the main text for
properties of the $\theta^k_k$'s.}}
$$
c^{(-1,q)} = d \left( \vartheta_0^0\cdot  c^{(0,q-1)} +
\vartheta^1_1\cdot  c^{(1,q-2)} + \cdots +
\vartheta^{q-1}_{q-1}\cdot c^{(q-1,0)} \right) +\vartheta^q_q
\cdot c^{(q,-1)}.
$$

The \emph{Collating Formula} is related to the Weil theorem which
can be rewritten neatly using the {\BD} machinery. \noindent

First, observe that $C^{p}_{p+1}=C_{p}^{p}$ but $Z^{p}_{p+1}
\subset Z_{p}^{p}$. Indeed on $\Om ^{(0,p-1)} \subset C^p _{p+1}$
the operator $\td$ is the genuine de Rham differential, while on
$\Om ^{(0,p-1)} \subset C^p _p$ it is the $0$ map, so
 the condition to be $D$-closed is more stringent in the first
case.
 If
$c=(c^{(0,p-1)},c^{(1,p-2)},\cdots,c^{(0,p-1)},\teger{c}{}^{(p,-1)})
$ belongs to $Z^{p}_{p}$, the standard de Rham differential
applied to $c^{(0,p-1)}$ leads to a global closed $p$-form.
Indeed, $\delta d c^{(0,p-1)}=d \delta c^{(0,p-1)}=\pm d^{\, 2}\,
c^{(1,p-2)}=0$,
 so  $d c^{(0,p-1)}$ is the restriction
of a global $p$-form, which is obviously closed. So there is a
canonical map $\left\{\mbox{Ker } D
     : C^{p}_{p}\rightarrow C_{p}^{p+1}\right\} \stackrel{m}{\longrightarrow}
     \left\{\mbox{Ker } d : \Om  ^{(-1,p)}
     \rightarrow \Om  ^{(-1,p+1)}\right\}$.
The image of this map is not totally obvious, but this is
precisely the content of  Weil's theorem \cite{AW52}: the sequence
of abelian groups
$$\xymatrix@1{
0 \ar[r] &Z_{p+1}^{p} \ar[r]^i &Z_{p}^{p} \ar[r]^-m
&{\left\{\scriptsize{\begin{array}{c}
\mbox{Closed global $p$-forms} \\
\mbox{with integral periods} \end{array}} \right\}} \ar[r] &0 }
$$
  is exact.

\subsubsection{Refinements}

If the simple covering $\mathcal{V}=
\{\mathcal{V}_{\sigma}\}_{\sigma \in
  J}$ of $M$ is a refinement of the simple covering $\mathcal{U}=
\{\mathcal{U}_{\alpha}\}_{\alpha \in I}$ and $\{\varphi_{\sigma}\}$
is a partition of unity for $\mathcal{V}$, we define a (compatible)
partition of unity for $\mathcal{U}$ $\{\th_{\alpha}\}=
\displaystyle \{\sum_{\sigma \in J \atop r(\sigma)=\alpha}
\varphi_{\sigma}\}$. For compatible partitions of unity, the
homotopy operator commutes with restriction, {\emph{i.e.}}  $K \circ
r=r\circ K$ (it being understood that the homotopy operator on the
left-hand side is for the covering $\mathcal{V}$ and on the
right-hand side for the covering $\mathcal{U}$). To summarize,
restriction commutes with $\delta$, $\td$, $D$ and $K$ :
\begin{equation} \label{eq:restrcom}
\delta \circ r=r\circ \delta \ , \ \td \circ r= r\circ \td \ , \ K
\circ r=r\circ K \ , \ D \circ r=r\circ D.
\end{equation}

We could say this more pedantically by drawing the \BD\ complexes
(or their enlarged versions) for $\mathcal{U}$ and $\mathcal{V}$
on top of each other (in three dimensions) and stating that
restriction is a (co-)chain map for all differentials or
co-differentials defined up to now.

\subsection{Computation of $H^{q}(\mathcal{C}_p,D)$, $q >
p$}{\label{hpplus}} \hspace{1.0cm}We show that for $q > p$ ,
\[ H^{q}(\mathcal{C}_p,D) \simeq H^{q}_{\check{C}ech}(M,\mathbb{Z}) \ \ \ \ \ \  (II)\]
(the isomorphism is canonical).

{\textbf{Proof:}} Start from the simple observation that for $q >
p$ one has the inclusion $K\td(C_p^q)\subset C_p^q$, so that one
can  freely use $(1+K\td)D=\delta(1+K\td)$ to compute
$H^{q}(\mathcal{C}_p,D)$.

The middle cohomology in the complex $\xymatrix@1{0 \ar[r]
&C_p^{q-1} \ar[r]^\delta &C_p^{q} \ar[r]^\delta &C_p^{q+1} \ar[r]
&0}$ is concentrated in de Rham degree $-1$ because $\delta$ does
not change the de Rham degree and has no cohomology in nonnegative
de Rham degree due to the existence of the homotopy operator. So
this cohomology is simply $H^{q}_{\check{C}ech}(M,\mathbb{Z})$. If
$\cg {}^{(q,-1)}\in \Om ^{(q,-1)}(\mathbb{Z})\subset C_p^{q}\,$ is a
\tch\ cocycle, $(1+K\td)^{-1}\,\cg {}^{(q,-1)}$ is a $D$-cocycle.
Conversely if the cochain $c=(c^{(q-p,p-1)}, \cdots, \\ \cg
{}^{(q,-1)})
 \in C_p^{q} $ is a $D$-cocycle, $\cg {}^{(q,-1)}$ is a \tch\ cocycle and
the relation $(*)$ is satisfied {\emph{i.e.}}
\[c=(1+K\td)^{-1}\,\cg {}^{(q,-1)} +D(1+K\td)^{-1}K (c-\cg {}^{(q,-1)}).\]
 Hence the projection map  $ \pi :C_p^q \rightarrow \Om
^{(q,-1)}(\mathbb{Z})$ descends to an isomorphism in cohomology
which proves the announced result $(II)$.

\subsection{The case $q = p$}{\label{hpegal}}

\hspace{1.0cm}A full description of $H^{p}(\mathcal{C}_p,D)$ is
complicated in general, but it fits in all cases into an exact
sequence of abelian groups\footnote{For instance, when $p=2$, we
recover the classification of line bundles with connection modulo
gauge equivalence, as expected. This case is treated in detail in
appendix \ref{app3}.}
\begin{eqnarray*}
\xymatrix@1{ 0 \ar[r] &{\left\{\scriptsize \begin{array}{c}
\mbox{Closed global $(p-1)$-forms} \\ \mbox{with integral periods}
\end{array} \right\}} \ar[r] &\Om ^{p-1} (M,\mathbb{R}) \ar[r]
&H^{p}(\mathcal{C}_p,D) \ar[r] &H^{p}_{\check{C}ech}(M,\mathbb{Z})
\ar[r] &0}  \\
  (III)
\end{eqnarray*}
{\textbf{Proof:}} We shall treat separately the cases $p = q = 0$
and $p = q \neq 0$, starting with the latter.

Let $c= (c^{(0,p-1)},  \cdots,c^{(p-1,0)},\cg{}^{(p,-1)}) \in
C_p^{p}\,$ be a $D$-cocycle, then $\cg {}^{(p,-1)}$ is a \tch\
cocycle and $(*)$ tells us that
\[c-(1+K\td)^{-1}\,\cg {}^{(p,-1)} =D(1+K\td)^{-1}K (c-\,\cg {}^{(p,-1)}).\]

\noindent However, we now have $K\td(C_p^q)\subset C_p^q + \Om
^{(-1,q)} (\mathbb{R})$, in contrast with the previous case for
which we had the inclusion $K\td(C_p^q)\subset C_p^q$.
Accordingly, as an element of $C_p ^{p-1} +\Om
^{(-1,p-1)}(\mathbb{R})$, $(1+K\td)^{-1}K (c-\, \cg {}^{(p,-1)})$
has a component, say $\gamma^{(-1,p-1)}$, in $\Om
^{(-1,p-1)}(\mathbb{R})$, so we cannot conclude that $c$ and
$(1+K\td)^{-1}\cg {}^{(p,-1)}$ are D-cohomologous. Nevertheless
$\td \gamma^{(-1,p-1)}=0$ (not $d$ !), hence $c$ is D-cohomologous
to $(1+K\td)^{-1}\,\cg {}^{(p,-1)}+\delta \gamma^{(-1,p-1)}$.

 Conversely, the cochain
$ (1+K\td)^{-1}\cg {}^{(p,-1)}+\delta \gamma^{(-1,p-1)}$ is a \BD\
cocycle whenever $\gamma^{(-1,p-1)}$ is a global de Rham
$(p-1)$-form and $\cg {}^{(p,-1)}\in \Om ^{(p,-1)}(\mathbb{Z})$ is
a \tch\ cocycle.

So we have exhibited a family of ``reduced" representatives
\[
(1+K\td)^{-1}\cg {}^{(p,-1)}+\delta \gamma^{(-1,p-1)}, \ \ \ \
(**)\] of \BD\ cohomology classes.

Decomposition $(**)$ leads us to consider the following maps  \[
\pi : c= (c^{(0,p-1)},  \cdots,c^{(p-1,0)},\cg{}^{(p,-1)}) \in
C_p^{p} \ \mapsto \ \cg{}^{(p,-1)} \in \Omega^{(p,-1)}
(M,\mathbb{Z}), \] (already met in subsection A.2), and \[ \phi :
\gamma^{(-1,p-1)} \in \Om ^{(-1,p-1)}(\mathbb{R}) \ \mapsto \
(\delta \gamma^{(-1,p-1)},0,\ldots,0 ) \in C_p^{p} .\]

\noindent We provide $\Om ^{(-1,p-1)}$ with the trivial
differential $ = 0$, so that $ \pi$ and $ \phi$ are maps between
complexes. It is quite easy to check that these two maps are chain
maps, \emph{i.e.} $ \phi \cdot 0 = D \cdot  \phi$ and $ \pi \cdot
D= \delta \cdot \pi$, hence, passing to cohomology,
$$
\xymatrix@1{ \Om ^{(-1,p-1)}(\mathbb{R}) \ar[r]^{\hat{\phi}}
&H^{p}(\mathcal{C}_p,D) \ar[r]^{\hat{\pi}}
&H^{p}_{\check{C}ech}(M,\mathbb{Z})}. $$
 Let us show that
$\hat{\pi}$ is surjective. First, by definition and with obvious
notations,
\[\hat{\pi}([c]):=[\cg {}^{(p,-1)}].
\] For any class  $\mathbf{\xi} \in
H^{p}_{\check{C}ech}(M,\mathbb{Z})$, let us pick a representative
$\cg {}^{(p,-1)}$ of $\mathbf{\xi}$. From $(**)$, we deduce that
\[ c = (1+K\td)^{-1} \cg {}^{(p,-1)} \] is a \BD\
cocycle which trivially fulfills $ \pi (c) =   \cg {}^{(p,-1)},$
so that \[\hat{\pi}([c])=[\cg{}^{(p,-1)}]=\mathbf{\xi}.\]  This
means that any integral \tch\ cohomology class is the image under
$\hat{\pi}$ of a \BD\ cohomology class, thus establishing the
surjectivity of $\hat{\pi}$.

According to this, we can extend  further the previous exact
sequence to the right
$$
\xymatrix@1{ \Om ^{(-1,p-1)}(\mathbb{R}) \ar[r]^{\hat{\phi}}
&H^{p}(\mathcal{C}_p,D) \ar[r]^{\hat{\pi}}
&H^{p}_{\check{C}ech}(M,\mathbb{Z})\ar[r] &0}. $$
 Now, let us
show that this sequence is actually exact on the left, that is to
say $Ker(\hat{\pi})= Im(\hat{\phi})$.

If $[c] \in Ker(\hat{\pi})$ then $\hat{\pi}([c]) = [\cg
{}^{(p,-1)}] = 0$, meaning that any representative of $[\cg
{}^{(p,-1)}]$ is a \tch\ coboundary, $\delta
\teger{\lambda}{}^{(p-1,-1)}$. Thus, if $(1+K\td)^{-1}\cg
{}^{(p,-1)} + \delta \gamma^{(-1,p-1)}$ is a ``reduced"
representative of $[c] \in Ker(\hat{\pi})$, we have
\[ c = (1+K\td)^{-1} \delta \teger{\lambda}{}^{(p-1,-1)} +
\delta \gamma^{(-1,p-1)} = Dq + \delta \rho^{(-1,p-1)},  \] with
$\rho^{(-1,p-1)} = \gamma^{(-1,p-1)} +
(-K\td)^{p}\teger{\lambda}{}^{(p-1,-1)}$. In other words,
$Ker(\hat{\pi})$ is made of \BD\ classes $[c]$ that admit a
representative of the form $\delta \rho^{(-1,p-1)}$ for some
global form $\rho^{(-1,p-1)} \in \Om ^{(-1,p-1)}(\mathbb{R})$.
Conversely, for any global form $\rho^{(-1,p-1)} \in \Om
^{(-1,p-1)}(\mathbb{R})$ the \BD\ class $[\delta \rho^{(-1,p-1)}]$
trivially belongs to $Ker(\hat{\pi})$.
 This implies $Ker(\hat{\pi})=Im(\hat{\phi})$.

So, the sequence

$$\xymatrix@1{ \Om ^{(-1,p-1)}(\mathbb{R}) \ar[r]^{\hat{\phi}}
&H^{p}(\mathcal{C}_p,D) \ar[r]^{\hat{\pi}}
&H^{p}_{\check{C}ech}(M,\mathbb{Z})\ar[r] &0}, $$

 is exact,
and to extend it to the left, we have to compute
$Ker(\hat{\phi})$.

If $\gamma^{(-1,p-1)} \in Ker({\hat{\phi}})$, then
${\hat{\phi}}(\gamma^{(-1,p-1)}) = [\delta \gamma^{(-1,p-1)}] =
0$, which means that any representative of $[\delta
\gamma^{(-1,p-1)}]$ is a \BD\ coboundary. In particular
\[\delta \gamma^{(-1,p-1)} = (\delta
\gamma^{(-1,p-1)},0, \ldots, 0 )= D\tau, \] for some
$\tau=(\tau^{(0,p-2)},
  \cdots ,\tau^{(p-2,0)}, \teger{\tau}
{}^{(p-1,-1)}) \in C_p^{p-1}$. This gives rise to the following
{\tch}-de Rham cochain
\[( -
\gamma^{(-1,p-1)}, \tau^{(0,p-2)}, \cdots
,\tau^{(p-2,0)},\teger{\tau} {}^{(p-1,-1)}),
\] which turns out to be a {\tch}-de Rham cocycle since $\delta
\gamma^{(-1,p-1)} = D\tau$. Now, from Weil's theorem (see subsection
A.4.3) we conclude that since $\teger{\tau} {}^{(p-1,-1)}$ is
integral the global form $\gamma^{(-1,p-1)}$ has integral periods.
Conversely, if $\gamma^{(-1,p-1)}$ has integral periods then, still
from Weil's theorem, it gives rise to an integral {\tch}-de Rham
cocycle $( \tau^{(-1,p-1)} = - \gamma^{(-1,p-1)}, \tau^{(0,p-2)},
\cdots ,\tau^{(p-2,0)},\teger{\tau} {}^{(p-1,-1)})$ such that
$\delta \gamma^{(-1,p-1)} = D\tau$. This shows that
$Ker(\hat{\phi})$ is nothing else but the space of $(p-1)$-forms
with integral periods. So we can extend our exact sequence to the
left using the canonical injection of $(p-1)$-forms with integral
periods into $(p-1)$-forms

$$\xymatrix@1{{\left\{\scriptsize \begin{array}{c} \mbox{Closed
global $(p-1)$-forms} \\ \mbox{with integral periods} \end{array}
\right\}} \ar[r]^-i &\Om ^{(-1,p-1)}(\mathbb{R})
\ar[r]^{\hat{\phi}} &H^{p}(\mathcal{C}_p,D) \ar[r]^{\hat{\pi}}
&H^{p}_{\check{C}ech}(M,\mathbb{Z}) \ar[r] &0}. $$

Finally, it is obvious that $Ker(i) = {0}$. This last point
definitively establishes the exactness of $(III)$ for $p = q \neq
0$. $\\$

In the special case $p = q = 0$, identity $(**)$ reads
\[
(1+K\td)^{-1}\cg {}^{(0,-1)}+ \delta \gamma^{(-1,-1)} = \cg
{}^{(0,-1)}+ \delta \gamma^{(-1,-1)}, \] where $\gamma^{(-1,-1)}$
is just a real number. This means that reduced representatives of
$[c] \in H^{0}(\mathcal{C}_0,D)$ are integral \tch\ cohomology
classes (canonically imbedded in the real \tch\ cohomology).
Conversely, any \tch\ cohomology class $\mathbf{\xi} \in
H^{0}_{\check{C}ech}(M,\mathbb{Z})$ defines a \BD\ class
$[(1+K\td)^{-1}\cg {}^{(0,-1)}]$, {\em{i.e.}}
$H^{0}(\mathcal{C}_0,D) \simeq
H^{0}_{\check{C}ech}(M,\mathbb{Z})$. As a side note, this can be
combined with $(II)$ to yield the more general result
\[H^{q}(\mathcal{C}_0,D) \simeq
H^{q}_{\check{C}ech}(M,\mathbb{Z}). \]

If $H^{p}_{\check{C}ech}(M,\mathbb{Z})$ has no torsion, the
sequence $(III)$ is split : choose a basis of
$H^{p}_{\check{C}ech}(M,\mathbb{Z})$, take a representative \tch\
cocycle in $\Om ^{p} (M,\mathbb{Z})$ for each basis element, and
multiply it by $(1+K\td)^{-1}$ to get a \BD\ cocycle, then extend
by linearity. This gives an injection of
$H^{p}_{\check{C}ech}(M,\mathbb{Z})$ into $H^{p}(\mathcal{C}_p,D)$
which is isomorphic (as an abelian group, but in a non canonical
way) to
\[ H^{p}_{\check{C}ech}(M,\mathbb{Z}) \oplus \Om  ^{p-1}
(M,\mathbb{R})/{\scriptsize \left\{\begin{array}{c} \mbox{Closed
global $(p-1)$
        forms} \\ \mbox{with integral periods} \end{array}
  \right\}}.\]

If $H^{p}_{\check{C}ech}(M,\mathbb{Z})$ has torsion there is no
splitting and the above description is not correct. Finally note
the special case $p=q=1$ : $H^1(\mathcal{C}_1,D)$ is canonically
isomorphic to $C^{\infty}(M,\mathbb{R}/\mathbb{Z})$, the
multiplicative group of smooth functions from $M$ to the circle
group, a more compact description than the one given by the exact
sequence $(III)$.

\subsection{The isomorphism between Cheeger-Simons differential
characters and {\BD} classes for $q=p$} \label{isoCS}

\hspace{1.0cm}The {\BD} cohomology group can be imbedded into
another exact sequence
$$
\xymatrix@1{ 0 \ar[r] &H^{p-1}_{\check{C}ech}(M,{\mathbb{R}}/
{\mathbb{Z}}) \ar[r] &H^{p}(\mathcal{C}_p,D) \ar[r] &\Om
^{p}_{\mathbb{Z}} (M,\mathbb{R}) \ar[r] &0} ,$$ which fits better
with the representation  we have chosen for the classes, namely:
 $$\om= (\om^{(0,p-1)},
\cdots,\om^{(p-1,0)},\teger{\om}{}^{(p,-1)})\,$$ On the other
hand, the Cheeger-Simons differential character group
$\hat{H}^{p}(M, {\mathbb{R}}/ {\mathbb{Z}})$ can also be imbedded
into the same exact sequence \cite{CS73,HLZ03}
$$
\xymatrix@1{ 0 \ar[r] &H^{p-1}_{\check{C}ech}(M,{\mathbb{R}}/
{\mathbb{Z}}) \ar[r] &\hat{H}^{p}(M, {\mathbb{R}}/ {\mathbb{Z}})
\ar[r] &\Om ^{p}_{\mathbb{Z}} (M,\mathbb{R}) \ar[r] &0} .$$

These two sequences can be combined into the following commutative
diagram

%\begin{equation}
$$
\xymatrix{
 0 \ar[r] &H^{p-1}_{\check{C}ech}(M,{\mathbb{R}}/
{\mathbb{Z}}) \ar[d] \ar[r] &H^{p}(\mathcal{C}_p,D)
\ar[d]^{\mbox{\large${\int}$}}
\ar[r] &\Om^{p}_{\mathbb{Z}}(M,\mathbb{R}) \ar[d] \ar[r]  &0 \\
0  \ar[r] &H^{p-1}_{\check{C}ech}(M,{\mathbb{R}}/{\mathbb{Z}})
\ar[u]^{\mbox{id}} \ar[r] &\hat{H}^{p}(M,
{\mathbb{R}}/{\mathbb{Z}}) \ar[r]
&\Omega^{p}_{\mathbb{Z}}(M,\mathbb{R}) \ar[u]^{\mbox{id}} \ar[r]
&0  }
$$
%\end{equation}
in which the descending map in the middle -$\int$- is given by
(\ref{integclassomega}). Then by the \emph{5-Lemma} this map is an
isomorphism.
\subsection{The \BD\ cohomology is the same for all good coverings}
\label{cover}

\hspace{1.0cm}A proof is needed only when $q=p$, because in the
other cases, we have given canonical isomorphisms with standard
\tch\ cohomology spaces.

If the simple covering $\mathcal{V}$ of $M$ is a refinement of the
simple covering $\mathcal{U}$, it is a classical theorem that for
\tch\ cohomology the restriction chain map induces an isomorphism
in cohomology.  This isomorphism is canonical because restriction
is canonical.

We use this as a starting point to prove the corresponding result
for \BD\ cohomology. To avoid notational ambiguities, we write
$\mathcal{C}_p(\mathcal{U})$ (resp.  $\mathcal{C}_p(\mathcal{V})$)
for the \BD\ complex for the covering $\mathcal{U}$ (resp.
$\mathcal{V}$).

Restriction gives a chain map from the complex
$(\mathcal{C}_p(\mathcal{U}),D)$ to the complex
$(\mathcal{C}_p(\mathcal{V}),D)$. So there is a canonical
homomorphism
\[H^{p}(\mathcal{C}(\mathcal{U})_p,D)
\stackrel{\scriptsize \mbox{restriction}}{\longrightarrow}
H^{p}(\mathcal{C}(\mathcal{V})_p,D).\]

We want to show that this homomorphism is one-to-one
onto\footnote{The general canonical isomorphism theorem for two
(arbitrary) simple coverings is an automatic consequence of the
fact that on a compact manifold two simple coverings have a common
simple refinement.}. We start by showing that the homomorphism is
one to one. Suppose that an element of
$H^{p}(\mathcal{C}(\mathcal{U})_p,D)$, represented by a certain
$c=(c^{(0,p-1)}, \cdots,c^{(p-1,0)},\cg {}^{(p,-1)} ) \in
\mathcal{C}(\mathcal{U})_p^{p}$, maps to the trivial element in
$H^{p}(\mathcal{C}(\mathcal{V})_p,D)$. This implies that the
restriction of $c^{(p,-1)}$ to the covering
$\{\mathcal{V}_{\sigma}\}_{\sigma \in J}$ is a trivial \tch\
cocycle, and by the isomorphism theorem for \tch\ cohomology,
$c^{(p,-1)}$ itself is trivial. From the previous section, we know
then that $c$ is \BD\ cohomologous to some $\delta \gamma
^{(p-1)}$ where $\gamma ^{(p)}$ is a global de Rham $(p-1)$ form,
so we can assume that $c=\delta \gamma ^{(p-1)}$ to start with.
The condition of triviality is then the same for both coverings,
{\ie} $\gamma ^{(p-1)}$ has to be closed with integral periods. We
have proved that in the diagram

$$\xymatrix{ 0 \ar[d] \\
H^{p}(\mathcal{C}(\mathcal{U})_p,D) \ar[r] \ar[d]^r &
H^{p}_{\check{C}ech}(M,\mathbb{Z})  \ar[r] \ar[d]^{\mathrm{Id}}  &0 \\
H^{p}(\mathcal{C}(\mathcal{V})_p,D)  \ar[r]
&H^{p}_{\check{C}ech}(M,\mathbb{Z}) \ar[r] &0 } $$ the first
column is exact ({\emph{i.e.}} restriction is one to one) and the
kernels of the top and bottom rows are canonically isomorphic via
restriction.

To prove that the restriction map is onto, we take compatible
partitions of unity $\{\th _{\alpha}\}$ and $\{\varphi_{\sigma}\}$
for $\mathcal{U}$ and its refinement $\mathcal{V}$. Take a class
in $H^{p}(\mathcal{C}(\mathcal{V})_p,D)$, represented by a cocycle
$s=(s^{(0,p-1)}, \cdots,s^{(p-1,0)},s^{(p,-1)})$ in
$\mathcal{C}(\mathcal{V})_p^{p}.$ Then $s^{(p,-1)}$ is a \tch\
cocycle for $\mathcal{V}$. If $s^{(p-1,-1)}$ is an integral \tch\
cochain of degree $(p-1)$ for $\mathcal{V}$,
$s+Ds^{(p-1,-1)}=(\cdots,s^{(p,-1)}+\delta s^{(p-1,-1)})$
represents the same \BD\ class, so by the isomorphism theorem for
\tch\ cohomology, we can assume without loss of generality that
$s^{(p,-1)}$ is the restriction of a \tch\ cocycle $c^{(p,-1)}$
for $\{\mathcal{U}_{\alpha}\}_{\alpha \in I}$. We have proved in
the previous section that $s$ is \BD\ cohomologous to
$(1+K\td)^{-1}s^{(p,-1)}+\delta
\gamma^{(p-1)}=(\cdots,s^{(p,-1)})$ for some global de Rham
$(p-1)$-form $\gamma^{(p-1)}$ (in this formula, $K$, $\td$ and
$\delta$ are with respect to the covering $\mathcal{V}$) so we can
assume without loss of generality that $s$ is of that form to
start with. Then $(1+K\td)^{-1}c^{(p,-1)}+\delta \gamma^{(p-1)}$
(where now $K$, $\td$ and $\delta$ are with respect to the
covering $\mathcal{U}$) is a \BD\ cocycle, and (as restriction
commutes with $K$, $\td$ and $\delta$), $s=r(c)$. So each element
of $H^{p}(\mathcal{C}(\mathcal{V})_p,D)$ has a representative
which is the restriction of a \BD\ $p$-cocycle for $\mathcal{U}$ :
the restriction chain map leads to a surjective map in \BD\
cohomology.
Putting things together, the proof that restriction
induces a canonical bijection from
$H^{p}(\mathcal{C}(\mathcal{U})_p,D)$ to
$H^{p}(\mathcal{C}(\mathcal{V})_p,D)$ is complete.

\sect{{\BD} dual of a cycle} \label{app4}
\hspace{1.0cm}In this
section we present a construction of a ``cycle map" which
associates a {\BD} cohomology class to a given cycle. The kind of
duality that is implied is not of the ``Poincaré" type, but is
rather an analog of Pontrjagin duality for {\BD} cohomology.

 Let $z_p$ be a {\em{singular}} or rather a {\emph {de Rham}}
 (cf. section (\ref{smog}))  integral
$p$-cycle of $M$ and $\mathcal{U}$ a simple cover. We perform the
following descent {\footnote{All chains involved below are
integral chains.}} using the singular boundary operator, $b$, and
 the {\tch} coboundary operator, $\delta$:
\begin{equation}\label{d1}
(\delta z_p)_{{\al}_0} =  z_{p \,|{\al}_0} = b\ {c^0 _{p+1}}
{}_{,\,{\al}_0} \ \ \ \mbox{in}\ \ \ \mathcal{U}_{{\al}_0}.
\end{equation}
Then
\begin{equation}
b\, ({c^0 _{p+1}} {}_{,\,{\al}_1} -{c^0 _{p+1}} {}_{,\,{\al}_0})
=z_{p\,|{\al}_1} - z_{p\,|{\al}_0}= 0 \ \ \ \mbox{in} \ \ \
\mathcal{U}_{{\al}_0 {\al}_1},
\end{equation}
so that
\begin{equation}
\lp \delta\, {c^0 _{p+1}} \rp_{{\al}_0 {\al}_1} := {c^0 _{p+1}}
{}_{,\,{\al}_1}-{c^0 _{p+1}} {}_{,\,{\al}_0}= b\, {c^1 _{p+2}}
{}_{,\,{\al}_0 {\al}_1} \ \ \ \mbox{in} \ \ \ \mathcal{U}_{{\al}_0
{\al}_1}.
\end{equation}
This descent goes on at level $k$ (the fact that the covering is
simple is  crucial):
\begin{equation} \delta \, {c^k _{p+k+1}} =
b\, {c^{k+1} _{p+k+2}} \end{equation} and stops for $k=n-p-2$
\begin{equation}\delta \, {c^{n-p-2} _{n-1}} = b\ {c^{n-p-1}
_{n}}.
\end{equation}
As usual ${c^{k+1} _{p+k+2}}$ is defined in
$\mathcal{U}_{{\alpha_0} \cdots {\alpha_{k+1}}}$.

Finally,
\begin{equation}
 \delta \, {c^{n-p-1} _{n}} =  {c^{n-p}_{n}} \ \ \ \mbox{with}\ \ \
 b\,
 {c^{n-p}_{n}}= 0,
\end{equation}
in each $\mathcal{U}_{{\alpha_0} \cdots {\alpha_{n-p}}}$. Hence
every ${c^{n-p}_{n}}{}_{,\,{\alpha_0} \cdots {\alpha_{n-p}}}$ is a
integral $n$-cycle in $\mathcal{U}_{{\alpha_0} \cdots
{\alpha_{n-p}}}$, so that we can write
\begin{equation}\label{cancocycle}
{c^{n-p}_{n}}{}_{,\,{\alpha_0} \cdots {\alpha_{n-p}}} = {\mathop
{\eta} \limits^{\mathbb Z}}_{z,\,{\alpha_0} \cdots {\alpha_{n-p}}}
\cdot \mathcal{U}_{{\alpha_0} \cdots {\alpha_{n-p}}},
\end{equation}
once $\mathcal{U}_{{\alpha_0} \cdots {\alpha_{n-p}}}$ has been
identified with a singular $n$-cycle in a natural way.
Furthermore, the ${\mathop {\eta} \limits^{\mathbb
Z}}_{z,\,{\alpha_0} \cdots {\alpha_{n-p}}}$'s define a {\tch}
cocycle in an obvious way. In terms of de Rham currents
\begin{equation}
\label{cur}  {c^k _{p+k+1}} \longrightarrow {{\eta}_z{}
^{(k,n-p-k-1)}},
\end{equation}
the above descent equations read
\begin{equation}\label{curdes}
\delta\, {{\eta}_z{} ^{(k,n-p-k-1)}}= d {{\eta}_z{}
^{(k+1,n-p-k-2)}}\ \ \ \ \cdots\ \ \ \   \delta\, {{\eta}_z{}
^{(n-p-2,1)}}= d {{\eta}_z{} ^{(n-p-1,0)}}.
\end{equation}
Now $$\delta\, {{\eta}_z{} ^{(n-p-1,0)}}=d_{-1} {{\eta}_z{}
^{(n-p,-1)}},$$  where one can show, using integration of
$n$-forms with compact supports in $\mathcal{U}_{{\alpha_0} \cdots
{\alpha_{n-p}}}$, that $${\eta}_z{}^{(n-p,-1)}_{{,\alpha_0} \cdots
{\alpha_{n-p}}}={\mathop {\eta} \limits^{\mathbb
Z}}_{z,\,{\alpha_0} \cdots {\alpha_{n-p}}}.$$ Therefore the
sequence $$\eta_{\mathcal{D}}^{(n-p-1)}(z) = (
{\eta}_z{}^{(0,n-p-1)},\ldots,{\eta}_z{}^{(n-p-1,0)} ,{\mathop
{\eta} \limits^{\mathbb Z}}_z)$$ fulfilling the descent
(\ref{curdes}) is nothing but a {\BD} cocycle with
{\em{distribution}} coefficients.

The singular homology that was used here (in the intersections of
the simple covering) is not the usual one (\emph{i.e.} with compact
support), but rather the ``infinite" one where chains may have
non-compact supports. Accordingly, the corresponding currents do not
necessarily have compact support in the intersections either.
Moreover, the {\tch} cocycle ${\mathop {\eta} \limits^{\mathbb
Z}}_z$ is \emph{a priori} non trivial since it is obtained from a
{\tch}-de Rham descent of the \emph{a priori} non trivial
integration current of $z$. In fact, ${\mathop {\eta}
\limits^{\mathbb Z}}_z$ is a {\tch} representative of the Poincaré
dual of $z$ on $M$.

Let us have a look at the ambiguities of the descent of the
$p$-cycle $z$ which led to $\eta_{\mathcal{D}}^{(n-p-1)}(z)$. At
the level of the currents ${\eta}_z{}^{(n-p-k,k-1)}$, one can
check that ambiguities of {\BD} type (\ref{Bambiguity}) are
obviously present. However, since our starting point is the
integral current of $z$, we could also have ambiguities on
${\eta}_z{}^{(0,n-p-1)}$ corresponding to the restriction of a
globally defined closed $(n-p-1)$-current, $\delta
{\eta}_z{}^{(-1,n-p-1)}$. But, since all the currents of our
descent \textbf{must} be integration currents of integral chains,
$\delta {\eta}_z{}^{(-1,n-p-1)}$ must necessarily be the
integration current of a $(p+1)$-cycle. Hence, it produces a {\BD}
ambiguity. The same argument holds at the bottom of the descent,
where our integral chains will only produce integral {\tch}
cochain ambiguities, which are also of {\BD} type.
In other words,
the fact we use integral chains to produce a {\BD} cocycle
provides us with a canonical {\BD} class
$[\eta_{\mathcal{D}}^{(n-p-1)}(z)]$ associated with $z$
\footnote{This result can be obtained using \emph{integrally flat}
currents defined in \cite{Fe96}, see also \cite{HLZ03}. }.

 \sect{$U(1)$ connections as
{\BD} cohomology classes} \label{app3}
\hspace{1.0cm}Let us
briefly recall how connections over $U(1)$-bundles are related to
{\BD} cohomology classes \cite{Bry93}. Let $P := P(M,U(1),E,\pi)$
be a principal $U(1)$-bundle with total space $E$ over $M$ and
projection $\pi$. For  a  given \textbf{simple} open covering of
$M$, ${{\mathcal{U}}}$, $P$ is described by  transition functions
$g_{\alpha \beta} : \mathcal{U}_{\al \beta} \mapsto U(1)$ which
 satisfy the cocycle condition
\begin{equation}\label{cocyleg}
g_{\alpha_0 \alpha_1}  g_{\alpha_1 \alpha_2}  g_{\alpha_2
\alpha_0} = 1 \ ,
\end{equation}
in any intersection $\mathcal{U}_{\alpha_0 \alpha_1 \alpha_2}$, or
equivalently
\begin{equation}\label{descenteF4}
 \Lambda_{\alpha_0 \alpha_1} +
\Lambda_{\alpha_1 \alpha_2} + \Lambda_{\alpha_2 \alpha_0} :=
n_{\alpha_0 \alpha_1 \alpha_2} \in \mathbb{Z}  \ ,
\end{equation} with
\begin{equation}\label{deflambda}
g_{\alpha_0 \alpha_1} = \exp(2 i \pi \Lambda_{\alpha_0 \alpha_1})
\end{equation}

Trivially
\begin{equation}\label{cocylen}
n_{\alpha_0 \alpha_1 \alpha_2} - n_{\alpha_0 \alpha_1 \alpha_3} +
n_{\alpha_0 \alpha_2 \alpha_3} - n_{\alpha_1 \alpha_2 \alpha_3} =
0 \ ,
\end{equation}
in $\mathcal{U}_{\alpha_0 \alpha_1 \alpha_2 \alpha_3}$, which
means that the collection $n^{(2,-1)}$ defined by these integers
is an {\bf integral {\tch} $2$-cocycle} on $M$.

Given a collection of local sections, a connection $\tilde{A}$ on
$P$ induces  a collection $(A)_\al $ of locally defined 1-forms on
$M$ which glue together on every $\mathcal{U}_{\alpha_0 \alpha_1}$
according to
\begin{equation}\label{glueA}
A_{\alpha_1} - A_{\alpha_0} = g^{-1}_{\alpha_0
\alpha_1}dg_{\alpha_0 \alpha_1}  \ =\ (2i\pi) \, d
\Lambda_{\alpha_0 \alpha_1} \ .
\end{equation}

We then obtain a family
\begin{equation}{\label{coc}}
(A^{(0,1)},\Lambda^{(1,0)},n^{(2,-1)}) \in
\check{C}^{(0)}({\cal{U}},\Omega^1(M))
 \times\check{C}^{(1)}({\cal{U}},\Omega^0(M)) \times
\check{C}^{(2)}({\cal{U}},\mathbb Z)\ ,\nonumber
\end{equation}
such that
\begin{eqnarray}\label{descenteF}
(\delta A^{(0,1)})_{\alpha_0 \alpha_1} &:=& A_{\alpha_1} -
A_{\alpha_0}
= (2i\pi) \, d \Lambda_{\alpha_0 \alpha_1} \ ,\\
(\delta \Lambda^{(1,0)})_{\alpha_0 \alpha_1 \alpha_2} &:=&
\Lambda_{\alpha_0 \alpha_1} + \Lambda_{\alpha_1 \alpha_2} +
\Lambda_{\alpha_2 \alpha_0} = d_{-1} n_{\alpha_0 \alpha_1
\alpha_2} := n_{\alpha_0
\alpha_1 \alpha_2} \ , \nonumber \\
(\delta n^{(2,-1)})_{\alpha_0 \alpha_1 \alpha_2 \alpha_3} &:=&
n_{\alpha_0 \alpha_1 \alpha_2} - n_{\alpha_0 \alpha_1 \alpha_3} +
n_{\alpha_0 \alpha_2 \alpha_3} - n_{\alpha_1 \alpha_2 \alpha_3} =
0 \nonumber \ ,
\end{eqnarray}
in the appropriate intersections. As described in detail above
such a sequence makes up a {\bf {\BD} cocycle}.

The curvature of $\tilde{A}$ also admits canonical local
representatives on $M$, $F_\alpha :=  d A_\alpha $, which are
globally defined since

\begin{equation}\label{descenteF1}
F_{\alpha_1} - F_{\alpha_0} = d (A_{\alpha_1} - A_{\alpha_0}) = 2i
\pi d (d \Lambda_{\alpha_0 \alpha_1}) = 0 \ ,
\end{equation}
 Obviously, the existence of $F$ on $M$ is a direct consequence
of the existence of $A^{(0,1)}$, and we can formally write ``$F =
d A^{(0,1)}$".

\vspace{5mm} For a given triple $( {\cal{U}}, P, \tilde{A})$ the
{\BD} cocycle $(A^{(0,1)},\Lambda^{(1,0)},n^{(2,-1)})$ is not
unique. More precisely, ambiguities on the local representatives
of $P$ and $\tilde{A}$ (that is allowed changes of transition
functions and local sections) induce ambiguities on the {\BD}
cocycle (\ref{coc}) of the following form
\begin{equation}\label{BDPA}
\left( d q^{(0,0)},  \delta q^{(0,0)} + d_{-1} m^{(1,-1)} ,\delta
m^{(1-,1)}
 \right), \end{equation}
 with $(m^{(1,-1)},q^{(0,0)}) \in
\check{C}^{(1)}({\cal{U}},\mathbb Z)\times
\check{C}^{(0)}({\cal{U}},\Omega^0(M))$ . Such ambiguities
correspond precisely to {\BD} coboundaries and thus represent the
ambiguities among the representatives of the relevant {\BD}
cohomology classes {\footnote{see appendix \ref{appDB}}}.

  Two triples $( {\cal{U}}, P, \tilde{A})$ and $( {\cal{U}}, P',
  \tilde{A'})$ are said to be \emph{$U(1)$-equivalent} if there is a $U(1)$
   isomorphism $\Phi :
P \mapsto P'$, such that $\tilde{A}' = {\Phi}_*\  \tilde A$.
Locally, this means that the transition functions of $P$ and $P'$
are related according to
\begin{equation}\label{equivg}
g'_{\alpha \beta} = {h_\alpha}^{-1}\cdot  g_{\alpha \beta}\cdot
h_\beta,
\end{equation}
or equivalently
\begin{equation}
 {\Lambda}'_{\alpha \beta} =\Lambda_{\alpha \beta} + q_\beta -
 q_\al,
\end{equation}
 where the $h_\al= \exp(2i\pi q_\al)$. In the same way the local
 representatives of the connections fulfill\begin{equation}
 {A'}_\al = A_\al + 2i\pi  dq_\al.
\end{equation}
Then we clearly see that these relations assume the same form as
the ambiguities in (\ref{BDPA}), showing that two equivalent
triples are associated to the same {\BD} cohomology class in
$H^2_{\mathcal{D}}(M,\mathbb{Z}(2))$.

 This correspondence can be
established in the reverse way. Indeed, consider a representative
$(A^{(0,1)},\Lambda^{(1,0)},n^{(2,-1)})$  of a {\BD} cohomology
class, the $U(1)$-valued mappings $g_{\alpha \beta} := \exp{2 i
\pi \Lambda_{\alpha \beta}}$ are $U(1)$ transition functions over
${\cal{U}}$ since they satisfy the cocycle condition
(\ref{cocyleg}). With these functions, one can canonically build a
principal $U(1)$-bundle over $M$, $P(M,U(1),E,\pi)
\cite{KN63}\cite{S74})$. Furthermore, there is only one connection
$\tilde{A}$ on $P$ whose local representatives on $M$ coincide
with those of $A^{(0,1)}$. Hence our {\BD} cocycle defines a
couple $(P,\tilde{A})$ in a canonical way.

Now, with another representative, $(A^{(0,1)} + d q^{(0,0)}  ,
\Lambda^{(1,0)} + \delta q^{(0,0)} ,n^{(2,-1)})$, we obtain
another set of transition functions which defines an equivalent
principal bundle -cf. (\ref{equivg}). In the same way, $A^{(0,1)}
+ d q^{(0,0)}$ is related to $\tilde{A}$ through a $U(1)$-bundle
isomorphism.

Finally, a representative $( A^{(0,1)}, \Lambda^{(1,0)} +
m^{(1,-1)} , n^{(2,-1)} + \delta m^{(1,-1)})$ gives the same
transition functions and the same connection. This establishes
that the {\BD} cohomology class of
$(A^{(0,1)},\Lambda^{(1,0)},n^{(2,-1)})$
 can be canonically associated to a whole class of
 $U(1)$-equivalent triples $( {\cal{U}}, P, \tilde{A})$.

\vspace{5mm} The independence of this isomorphism upon the chosen
covering  ${\cal{U}}$ of $M$ is a direct consequence of the
results proven in (\ref{cover})

\vspace{5mm} {\bf {Note added}}

 While completing this paper, we became aware of the
recent mathematical work of R. Harvey, B. Lawson and J. Zweck
\cite{HLZ03}, who discuss
 in detail the Pontrjagin duality we use
in Section (\ref{subsec:longf}). The authors emphasize the
differential character point of view rather than the {\BD} one we
have adopted here.

 \vspace{5mm} {\bf
{Acknowledgments}}

 R. S. wishes to thank the Erwin Schrödinger International Institute
 for Mathematical Physics in Vienna for the kind hospitality
 extended to him in 1998, at an early stage of this work, I. M.
 Singer for communicating an early version of \cite{Hop02}, and R.
 Zucchini for interesting discussions.

\end{document}